\documentstyle[pra,aps,multicol,psfig]{revtex}
\begin{document}


\draft

\title{The Superfluid State of Atomic $^6$Li in a Magnetic Trap}

\author{M. Houbiers, R. Ferwerda, and H.T.C. Stoof} 
\address{University of Utrecht, Institute for Theoretical
         Physics, P.O. Box 80.006, 3508 TA  Utrecht, The Netherlands}
\author{W.I. McAlexander, C.A. Sackett, and R.G. Hulet}
\address{Physics Department and Rice Quantum Institute,
         Rice University, Houston, Texas 77005} 
        
\maketitle

\begin{abstract}
We report on a study of the superfluid state of 
spin-polarized atomic $^6$Li confined in a magnetic trap.  
Density profiles of this degenerate Fermi gas, and the spatial distribution 
of the BCS order parameter are calculated in the local density
approximation. The critical temperature is determined as a function 
of the number of particles in the trap. Furthermore we 
consider the mechanical stability of an interacting two-component Fermi gas,
both in the case of attractive and repulsive interatomic interactions. 
For spin-polarized $^6$Li we also calculate the decay rate
of the gas, and show that within the mechanically stable regime of 
phase space, the lifetime is long enough to perform experiments on
the gas below and above the critical temperature if a bias
magnetic field of about 5 T is applied. Moreover, we propose that 
a measurement of the decay rate of the system might signal
the presence of the superfluid state. 
\end{abstract}

\pacs{PACS numbers: 03.75.Fi, 67.40.-w, 32.80.Pj, 42.50.Vk}
 
\begin{multicols}{2}

\section{INTRODUCTION}
\label{inleiding}
One of the most important objectives in the study of dilute gases, has been
the achievement of Bose-Einstein condensation (BEC) in bosonic systems.
Indeed, decades of experimental research finally lead two years ago to the 
observation of BEC in three different systems of alkali metal gases $^{87}$Rb,
$^7$Li and $^{23}$Na \cite{JILA,rice,MIT}. 
This success has triggered a large amount of interest in the field of
ultra-cold atomic gases. Although the study of properties of
these degenerate atomic Bose gases is vigorously being pursued at the moment, 
trapping and cooling of Fermi gases might also provide new and 
exciting physics. Indeed, in a previous theoretical study we
showed that a gas of spin-polarized atomic $^6$Li becomes superfluid
at densities and temperatures comparable with those at which the
Bose-Einstein experiments are performed \cite{henk}. 

This superfluid phase transition, which is similar to the BCS
transition in a superconductor, occurs at such high temperatures due to the 
fact that $^6$Li has an anomalously large and negative (triplet) $s$-wave 
scattering length $a$ \cite{randy2}. This scattering
length is a measure for the interatomic interactions, and its sign 
implies that this interaction is effectively attractive, which is a 
first requirement for a BCS transition to occur. 
For other atomic species, the transition temperature is in general
very low, because of the fact that the scattering length is of the order of
the range of the interaction $r_V$\ and the diluteness of the gas requires
that the Fermi wavenumber $k_F \ll 1/r_V$.
So for example in the case of deuterium, it was concluded already
some time ago that the observation
of a BCS transition is experimentally impossible \cite{leggett}.

\begin{figure}[htbp]
\psfig{figure=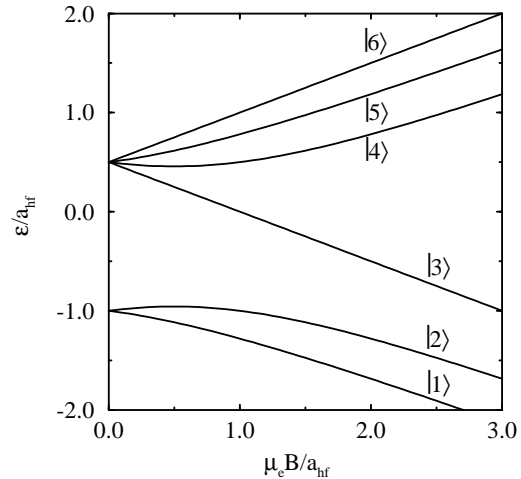}
\caption{\narrowtext
Energy of the six hyperfine states of $^6$Li
in units of the hyperfine constant $a_{hf}$, as a function of the magnetic field.}   
\label{fig1}
\end{figure}

The $^6$Li atom has nuclear spin $i=1$, and electron spin $s=1/2$.
Consequently the atom has six hyperfine states $|1\rangle$\ to $|6\rangle$, for
which the level splitting in a magnetic field is shown in Fig.~\ref{fig1}. 
The upper three levels $|4\rangle$\ to $|6\rangle$\ can be trapped in a 
static magnetic trap, whereas the lowest three hyperfine levels prefer 
high magnetic fields and are expelled from a magnetic field minimum.

The simplest way to create a degenerate Fermi gas is to trap just one 
low-field seeking hyperfine state, and for the sake of stability of the gas,
the doubly polarized state $|6\rangle=|m_s=1/2,m_i=1\rangle$\ is 
most suitable. However, due to the Pauli exclusion principle, two fermions in 
the same hyperfine state can interact with each other
at best via $p$-waves, and if this interaction is effectively attractive, 
the onset of the formation of Cooper pairs occurs at a temperature 
\[
T_c \simeq \frac{\epsilon_F}{k_B} \exp{\left\{ -\frac{\pi}{2(k_F |a|)^3} \right\}}
\]
where $\epsilon_F = \hbar^2k_F^2/2m$\ is the Fermi energy 
of the atomic gas, and $a$ the $p$-wave 
scattering length. For $^6$Li this $p$-wave scattering length
of the triplet potential is approximately $-35 a_0$, where $a_0$ is
the Bohr radius, 
and $k_Fa \ll 1$ in general. As a result the critical temperature 
for such a doubly spin polarized $^6$Li gas is extremely low. At present, 
a reasonable number for the density of trapped atomic gases is 
$10^{12}$cm$^{-3}$, leading to $\epsilon_F/k_B \simeq 600$nK,
and $k_F|a| \simeq 7 \times 10^{-3}$. The corresponding critical temperature
is clearly unattainable. 

In the case that more than one state is trapped,
Baranov {\it et al.} \cite{baranov} predicted a considerable increase in the
above ($p$-wave) critical temperature as a result of the fact that
two atoms in the same spin state can now also attract 
each other through the exchange of a phonon (density fluctuation)
in another hyperfine state. It was found that in this case the 
transition temperature
\[
T_c \simeq \frac{\epsilon_F}{k_B} \exp{ \left\{
-13\left(\frac{\pi}{2k_F |a|}\right)^2 \right\} },
\]
where $a$ now corresponds to the $s$-wave scattering length for collisions
between the two hyperfine states.
Nevertheless, using again a density of $10^{12}$ cm$^{-3}$ for each spin state
and the value $a=-2160a_0$ for $^6$Li \cite{randy2}, 
we find that $k_F|a| \simeq 0.43$,  
and it is easily verified that also in this case the critical temperature
is out of reach experimentally. 

Therefore, the most promising approach is to consider a 
Cooper pair of two atoms in different hyperfine states, since then
$s$-wave pairing is allowed. In this case \cite{gorkov}
\[
T_c \simeq \frac{\epsilon_F}{k_B} \exp{ \left\{ -\frac{\pi}{2k_F |a|} \right\} },
\]
resulting in a much higher critical temperature than in the
previous two cases. In particular, we envision to trap $^6$Li atoms 
in the states $|6\rangle$ and $|5\rangle$.
Experimentally, this might be achieved most easily by first trapping only
one hyperfine level, and then applying a {\it noisy} rf-pulse to
create an incoherent mixture of atoms occupying these two hyperfine levels 
\cite{peter}. Note that this situation has in fact already been realized 
in recent experiments with $^{87}$Rb atoms, although using a different
technique \cite{JILA2}.
 
In a recent publication Modawi and Leggett propose to trap $^6$Li atoms in
three instead of two hyperfine states \cite{tony2}. The advantage 
in such a system is that the effect of fluctuations is reduced
somewhat, but the disadvantage of trapping more hyperfine states, is
that the number of channels by which the gas can decay increases considerably.
There are not only more possibilities for two-body collision processes, in which
one or two electron spins are flipped and the corresponding atoms are
expelled from the trap, but also
three-body recombination processes are now no longer suppressed.
Therefore, at present, it seems to be most favorable to trap only
two hyperfine states, and the most suitable
candidates are the states $|6\rangle =|m_s=1/2;m_i=1
\rangle$, and $|5\rangle \simeq |m_s=1/2;m_i=0\rangle$, because for 
this combination the decay processes due to two-body interactions can be 
suppressed most. The approximate sign in the last expression indicates 
that in the state $|5\rangle$ there is for $\mu_e B \gg a_{hf}$
a small admixture with the spin state $|m_s=-1/2;m_i=1\rangle$. 
Although this admixture can be neglected for most purposes, we come back
to its importance for the stability of the gas shortly.

As explained above, in a two-component spin-polarized atomic $^6$Li gas, 
Cooper pairing will occur only between atoms in different
spinstates, while there is almost no interaction between two
atoms in the same spin states. For notational simplicity, we also 
refer to these states as $|\uparrow\rangle$ and $|\downarrow\rangle$,
and the densities
of atoms in these two hyperfine states are denoted by $n_{\uparrow}$ and
$n_{\downarrow}$, respectively. Notice that since the two states are
electron spin-polarized, the strength of the interatomic
interaction is indeed characterized by the $s$-wave scattering length of the
triplet potential $V_T({\bf r})$, and it is exactly this number which
is anomalously large and negative in the case of $^6$Li.  

The aim of the present publication is threefold. First,
the homogeneous calculation of Ref.~\cite{henk} needs some improvement,
due to the fact that the interatomic interaction potential for
$^6$Li has recently been determined more accurately \cite{randy2}.
The most up-to-date value of the $s$-wave scattering length is
$a=-2160 a_0$, where $a_0$\ is the Bohr radius. This change in
$a$\ not only affects the critical temperature but also the decay rates of the
gas. Second, we want to take the effect of the inhomogeneity of the
trapped gas into account, and in the third place, we    
look for a signature that signals the presence of the superfluid phase
in the gas.

The paper is organized as follows. In Sec.~\ref{rates}, we consider in some
detail the decay processes limiting the lifetime of the gas. 
Subsequently, we briefly summarize the theory for the
homogeneous Fermi gas in Sec.~\ref{BCStransition}\ and
improve the results obtained earlier for the critical temperature,
using the most up-to-date interatomic potential for $^6$Li. 
In Sec.~\ref{stability}\ we consider the mechanical 
stability of a weakly interacting Fermi gas. In particular we 
also consider a gas with positive $s$-wave scattering length, and 
show that in the unstable part of the phase diagram, 
a spinodal decomposition can restore the stability of the gas in this case. 

In future experiments the atoms are likely to be trapped in an external
potential that roughly has the shape of an isotropic
harmonic oscillator $V({\bf r}) = \frac{1}{2} m\omega^2 {\bf r}^2$, 
and which causes the gas cloud to be inhomogeneous. Therefore, 
the last part of this paper is devoted to the study of an 
inhomogeneous two-component Fermi gas at and below the critical temperature, and
in particular we will again concentrate on $^6$Li. 
The numerical calculations will be performed in the local density 
approximation, which is valid if the correlation length $\xi$\ over which the
particles influence each other is much smaller than the typical trap size $l =
\sqrt{\hbar /m \omega}$\ over which the density of the gas changes. 
A similar calculation for the noninteracting case has been performed recently
by Butts and Rokhsar \cite{rokhsar}. 
In addition, the case of purely repulsive interactions has been studied by
Oliva in the same way in the context of possible experiments with spin-polarized
atomic deuterium \cite{oliva}. 
In Sec.~\ref{LDA}, we briefly repeat the ingredients for the local
density approximation. In Sec.~\ref{Tc} we 
calculate the critical temperature of the gas as a function of the
number of trapped atoms, and in Sec.~\ref{benedenTc} we study
the gas in the superfluid state. Density profiles for the gas as well as for
the BCS order parameter are presented. 
In Sec.~\ref{conclusie}\ we devote a discussion to the issue of how
to detect the superfluid phase and to distinguish it from the normal phase.
We end the paper with a summary of the main conclusions. 

\section{HOMOGENEOUS FERMI GAS}
\label{homogeen}

We first consider a homogeneous, 
dilute gas of (electron) spin-polarized $^6$Li atoms.
Since the gas is dilute, the atoms will interact with each other mainly through
two-body collisions. These two-body collisions can be represented
on the mean-field level
by a local potential with a strength given by the two-body scattering
matrix $T^{2B}= 4 \pi a \hbar^2/m$, where $m$\ is the mass
of the particles and $a$\ is the scattering length \cite{henk2}. The
sign of $a$ determines whether the two-body interaction is effectively
repulsive ($a>0$), or attractive ($a<0$). 

Before going to a description of the gas in the normal and superfluid state, we 
consider an aspect that is experimentally of some 
importance, namely the lifetime
of the gas. The large $s$-wave scattering length has on the one hand the
advantage of having many thermalizing collisions between the particles which
is required for efficient evaporative cooling, but on the
other hand there will also be relatively many inelastic collisions which can cause
spin-flips within the atoms. If the electron spin of an atom is inverted, the atom
will be lost from the trap, and consequently these inelastic processes
limit the lifetime of the gas. In the next subsection we explain in more
detail which decay processes dominate in a mixture of $^6$Li atoms
in the hyperfine states $|6\rangle$ and $|5\rangle$.

\subsection{Decay rates}
\label{rates}
The electron spin and nuclear spin quantum numbers of the two trapped
hyperfine levels for $\mu_e B \gg a_{hf}$ are given by
\begin{eqnarray*}
|6\rangle & = & |m_s=1/2;m_i=1\rangle  \\
|5\rangle & = & |m_s=1/2;m_i=0\rangle + \theta^+
|m_s=-1/2;m_i=1\rangle, 
\end{eqnarray*}
where $\theta^+ \simeq a_{hf}/(2\sqrt{2} \mu_e B )$ 
is inversely proportional to the applied magnetic field $B$, so for
sufficiently strong magnetic fields the admixture of $|5 \rangle$
with the high-field seeking part is small and the gas can considered
to be electron spin-polarized. 
For such large magnetic fields, the energies of 
these two hyperfine levels are given by $\epsilon_{6}
= a_{hf}/2 + \mu_e B$ and $\epsilon_{5} \simeq \mu_e B$, respectively.

Since the two atoms in state $|5\rangle$ and $|6\rangle$ will interact at
the low temperatures of interest solely via
$s$-wave scattering, implying that the spatial part of the two-body 
wave function is symmetric under the exchange of atoms, 
the spin part of the wave function must be anti-symmetric, i.e.
\begin{eqnarray}
|\{6,5\}_-\rangle & = & \frac{1}{\sqrt{2}} \left[ |6\rangle|5\rangle
-|5\rangle|6\rangle \right] \nonumber \\
& = & |11;11\rangle + \theta^+|00;22\rangle,
\label{totalspin}
\end{eqnarray}
where in the last line we used the basis 
$|SM_S;IM_I\rangle$ with ${\bf S} = {\bf s}_1+ {\bf s}_2$ and
${\bf I}={\bf i}_1+{\bf i}_2$ the total electron and nuclear spin of 
the two colliding atoms, and $M_S$ and $M_I$ the corresponding magnetic
quantum numbers along the direction of the magnetic field.

The decay rates for the transition from the state $|lm,\{\alpha,\beta\}\rangle$ 
with orbital quantum numbers $l$ and $m$ to a state $|l'm',
\{\alpha',\beta'\}\rangle$ 
with quantum numbers $l'$ and $m'$, is essentially given by 
Fermi's Golden Rule and results in the expression \cite{henk4}
\begin{eqnarray}
G_{\alpha,\beta\rightarrow\alpha',\beta'}(B) & = & 2 \pi^3 \hbar^2 m p_{\alpha',
\beta'} \times \nonumber \\
& & \hspace{0.5cm}
\left| T_{l'm'\{\alpha',\beta'\},lm\{\alpha,\beta\}}(p_{\alpha',\beta'},0)
\right|^2,
\label{generalrate}
\end{eqnarray}
for the zero-temperature limit of the rate constant for this process.
Here $T_{l'm'\{\alpha',\beta'\},lm\{\alpha,\beta\}}(p_{\alpha',\beta'},0)$ is the
two-body scattering matrix at zero energy such that the incoming particles
have zero relative momentum, and the magnitude 
of the relative momentum of the scattered particles is $p_{\alpha',\beta'}$.
 
\begin{figure}[htbp]
\psfig{figure=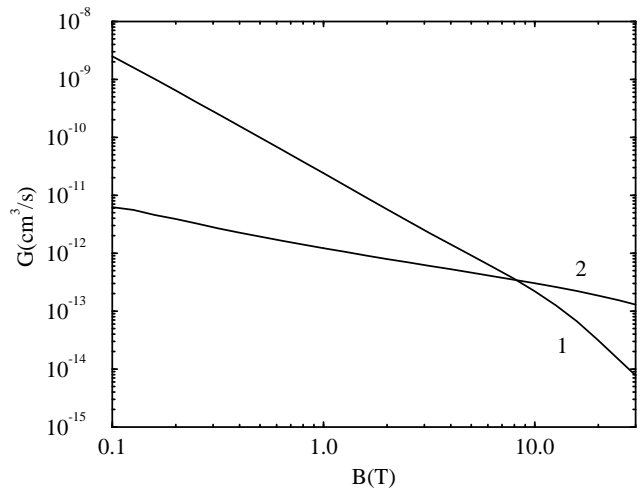}
\caption{\narrowtext
Decay rate constants due to exchange (curve 1) and one spin-flip 
processes (curve 2) as a function of the applied magnetic bias field.}
\label{fig2}
\end{figure}

As was explained for example in Ref.~\cite{henk}, there are basically two ways in
which collisions cause the atoms to be lost from the trap. First of all,
the central (singlet and triplet) interaction $V^c = V_S({\bf r})
{\cal P}^{(S)} + V_T({\bf r}) {\cal P}^{(T)}$\ induces
transitions between different hyperfine levels. Since this interaction
cannot change the total electron or nuclear spin angular momentum,
and the hyperfine level $|5\rangle$\ has a small admixture 
with the state $|m_s=-1/2,m_i=1\rangle$,
only transitions $|\{6,5\}_-\rangle \rightarrow |\{6,1\}_-\rangle$, where
$|1\rangle \simeq |m_s=-1/2,m_i=1\rangle - \theta^+|m_s=1/2,
m_i=0\rangle$\ are allowed. Similar to Eq.~(\ref{totalspin}), the
total spin state $|\{6,1\}_-\rangle $ is given by
\begin{equation}
|\{6,1\}_-\rangle = |00;22\rangle - \theta^+ |11;11\rangle.
\label{totalspin61}
\end{equation}
Combining Eqs.~(\ref{totalspin}) and (\ref{totalspin61}), we find
that the spin part of the transition matrix $T_{00\{6,1\},00\{6,5\}}(p_{61},0)$\ 
contributes a factor $\theta^+$ times the
exchange potential $V^{ex}({\bf r}) = V_T({\bf r}) - V_S({\bf r})$, i.e.~the 
difference between the triplet and singlet potential.  
To calculate the spatial part, we must use for the relative
in- and outcoming scattering wave functions with orbital quantum numbers $l$ and
$m$ and total electron spin $S$ the normalized expression
\begin{equation}
\Psi^{(\pm)}_{lmS}({\bf r}) = \sqrt{\frac{2}{\pi \hbar^3}} \frac{
\psi^{(\pm)}_{lS}(r)}{r} i^l Y_{lm}({\bf \hat{r}}),
\label{wavefunction}
\end{equation}   
where $\psi^{(\pm)}_{lS}(r)$ denotes the in- and outcoming solutions
to the radial Schr\"odinger equation with the singlet or triplet interaction. 
Using furthermore that the relative momentum $p_{61}$ after scattering
is due to the energy difference $\epsilon_6 - \epsilon_1
= 2 \mu_e B$, we find that $p_{61}=\sqrt{2m\mu_eB}$.
Combining all expressions into Eq.~(\ref{generalrate}), we obtain that the
rate constant due to exchange interactions is given by
\begin{eqnarray}
G^{ex} &  = & 2 \pi^3\hbar^2 m p_{61} (\theta^+)^2 \times \nonumber \\
& & \hspace{0.5cm}
\left| \langle \Psi^{(-)}_{000}({\bf r},p_{61})|V_T({\bf r}) - V_S({\bf r})|
\Psi^{(+)}_{001}({\bf r},0) \rangle \right|^2 \nonumber \\
 &  = & \pi^3\hbar^2 \left( \frac{m}{2\mu_e B}\right)^{\frac{3}{2}} a^2_{hf} 
 \times \nonumber \\
& & \hspace{0.5cm}
\left| \langle \Psi^{(-)}_{000}({\bf r},p_{61})|V^{ex}({\bf r})|
\Psi^{(+)}_{001}({\bf r},0) \rangle \right|^2.
\label{exchangerate}
\end{eqnarray}
In Fig.~\ref{fig2} this exchange rate as a function of the magnetic field
is shown (curve 1).

The second way in which collisions cause decay of the gas is due to
magnetic dipolar interactions $V^d$. As will be shown, of the various dipolar
interactions, the contribution due to electron-electron dipolar interactions
is most important. For this dipolar interaction, we have
\cite{henk4}
\begin{equation}
V^d = -\frac{\mu_0\mu_e^2}{4\pi r^3} \sqrt{\frac{4\pi}{5}}
\sum_{\Delta M_S} (-1)^{\Delta M_S} Y_{2 -\Delta M_S}({\bf \hat{r}}) 
\Sigma^{ee}_{2,\Delta M_S}, 
\label{dipolar}
\end{equation}
where the tensor operator $\Sigma^{ee}_{2,\Delta M_S}$ can be thought
of as arising from the coupling between ${\bf s}_{1}/\hbar$ and
${\bf s}_2/\hbar$, the Pauli spin matrices describing the electron spin 
of the two atoms, to a tensor of rank 2.  
For the scattering state $|\{6,5\}_-\rangle \simeq |11;11\rangle$, 
the dipolar interaction can change the (total) electron spin projection
$M_S$\ of the two atoms by an amount $\Delta M_S = -1$ for a one spin-flip
(1sf), or $\Delta M_S =-2$ for a two spin-flip (2sf) process.
Therefore, the one (two) spin-flip dipolar interaction couples 
the incoming wave function with approximately $S=1,M_S=1$ to the final state
having $S=1,M_S=0\ (M_S=-1)$. As a result, the outgoing wave
function is in the state $|10;11\rangle$ for one spin-flip, and in the 
total spin-state $|1-1;11\rangle$ after the two spin-flip interaction.
The Clebsch-Gordon coefficients
for each process are given by $\sqrt{3/10}$ and $\sqrt{3/5}$, i.e.~the
spin part of the transition matrix contributing to the decay rate 
is a factor of $\sqrt{2}$ larger for the two spin-flip
process than for the one spin-flip process.
Moreover, the energy released in a one spin-flip process is only half of
the energy released in a two spin-flip process. Therefore we find that
$p_{\alpha',\beta'}^{1sf} = \sqrt{2m\mu_eB}$\ whereas $p_{\alpha',\beta'}^{2sf}
= \sqrt{4 m \mu_e B}$. We thus arrive at the convenient relation
that $G^{2sf}(B) = 2G^{1sf}(2B)$\ and it suffices to calculate only
the one spin-flip decay rate.

Performing a similar calculation as in the case of the exchange decay rates,
the one spin-flip decay rate becomes
\begin{eqnarray}
G^{1sf}(B) & = & 2 \pi^3 \hbar^2 m \sqrt{2 m \mu_e B} \left|
\frac{\mu_0 \mu_e^2}{4 \pi} \sqrt{\frac{4 \pi}{5}}  \times \right. \nonumber \\
& & \hspace{-1cm} 
\langle \Psi^{(-)}_{211}({\bf r})| \left. \frac{Y_{2 1}({\bf \hat{r}})}{r^3} |
\Psi^{(+)}_{001}({\bf r}) \rangle \langle 10;11|\Sigma^{ee}_{2,-1}| 11;11\rangle
\right|^2 \nonumber \\
& = & \frac{12}{10} \sqrt{2 m \mu_e B} \frac{m(\mu_0 \mu_e^2)^2}{\pi \hbar^4}
(r_{20})^2,
\label{rate1sf}
\end{eqnarray}
where 
\[
(r_{20})^2 = \int_0^\infty dr \frac{\psi_{21}^{(-)}(r) \psi_{01}^{(+)}(r)}{r^3}
\]
is the radial electron-electron dipolar element.
In Fig.~\ref{fig2}, the one spin-flip decay rate constant is shown as curve 2.  

At this point it can be understood that the electron-electron dipolar
interaction gives the largest contribution to the dipolar decay rates. Decay
due to the electron-nucleon interaction occurs for example via the $|\{6,5\}_-\rangle
\rightarrow |\{6,4\}_-\rangle$-channel. However, the corresponding decay rates
are smaller by a factor of $(\mu_N/\mu_e)^2 \simeq 20 \times 10^{-6}$ and thus
completely negligible. This also implies that a mixture of $|6\rangle$ and 
$|5\rangle$ atoms cannot achieve equilibrium in the spin degrees of freedom
within the lifetime of the gas. This is completely analogous to the recent experiments
with two condensates in different spinstates  performed by the JILA group 
\cite{JILA2}.

Fig.~\ref{fig2} shows that the lifetime of the gas is of the
order of 1 s for a density $n_{5}=n_{6} \simeq 10^{12}$ cm$^{-3}$, 
and a magnetic bias field of 5 T. Although this would provide ample
time to perform an experiment, a much shorter lifetime may be adequate. 
For successful experiments we
not only have to require that the time between thermalizing collisions
is small compared to the lifetime of the gas, but also that the time scale
for formation of the Cooper pairs obeys this condition. The latter is 
anticipated to be of ${\cal O}(\hbar/k_BT_c)$ and therefore in our case 
much longer than the time between collisions. Nevertheless, for a density
$n_5=n_6 \simeq 10^{12}$ cm$^{-3}$, we have that $T_c \simeq 11$ nK,
and $\hbar/k_B T_c$ is only about $0.7$ ms. Our calculation overestimates the
spin-exchange rate constant below a magnetic field of 1 T,
and we estimate that a bias field of $0.2$ T would give a lifetime
of at least 1 ms. 

In the next subsection we consider the microscopic theory that describes the
Fermi gas in the normal and the superfluid state. We only
apply the BCS theory here. The influence of fluctuations \cite{gorkov}
will be addressed in a future publication.

\subsection{BCS transition}
\label{BCStransition}
For the homogeneous case, and taking only two-body interactions between
atoms in different hyperfine states into account,
the gas is described by the following hamiltonian \cite{FenW} 
\[
H  =  \sum_{\alpha=\uparrow,\downarrow} \int  d{\bf x}~ 
      \psi_{\alpha}^{\dagger}({\bf x})
      \left(- \frac{\hbar^2 \nabla^2}{2m} - \mu_{\alpha} \right)
      \psi_{\alpha}({\bf x})  
\]
\begin{equation}
  + \frac{1}{2} \int d{\bf x}\int d{\bf x}'~ V_T({\bf x}-{\bf x}') 
       \psi_{\alpha}^{\dagger}({\bf x})
              \psi_{-\alpha}^{\dagger}({\bf x}')
    \psi_{-\alpha}({\bf x}')\psi_{\alpha}({\bf x}). 
\label{H}
\end{equation}
In this expression, $\uparrow$\ and $\downarrow$\ refer again to the 
two hyperfine states involved. The field operators $\psi_{\alpha}(
{\bf x})$\ and $\psi_{\alpha}^{\dag}({\bf x})$\ obey the usual Fermi
anti-commutation relations, and denote the annihilation and creation
of a fermion at position ${\bf x}$\ in hyperfine state $|\alpha \rangle$\ with
chemical potential $\mu_{\alpha}$. The interparticle potential
can be approximated by a local potential, $V_T({\bf x}-{\bf x}') \simeq 
V_0 \delta({\bf x}-{\bf x}')$, where the constant $V_0$\ is a measure
of the strength of the interaction. We come back to the precise
value of $V_0$ shortly, but it is in any case negative to account for the effectively
attractive nature of the triplet interaction. The integration over ${\bf x}'$ 
in the hamiltonian is then trivial. The next step in a mean-field treatment of the 
hamiltonian in Eq.~(\ref{H}), is to develop the operator products 
$\psi_{\alpha}^{\dag}\psi_{\alpha}$ and $\psi_{\alpha} \psi_{-\alpha}$ around  
their mean values by substituting 
\[ \psi_{\alpha}^{\dag}\psi_{\alpha} = \langle \psi_{\alpha}^{\dag}
\psi_{\alpha} \rangle + \delta \psi_{\alpha}^{\dag} \psi_{\alpha}  
\]
and
\[ \psi_{-\alpha}\psi_{\alpha} = \langle \psi_{-\alpha}
\psi_{\alpha} \rangle + \delta \psi_{-\alpha} \psi_{\alpha}. 
\]
To first order in the fluctuations, we are left with the following
effective mean-field hamiltonian 
\begin{eqnarray}
H  =  \int & d\vec{x}~ & \left\{ \sum_{\alpha=\uparrow,\downarrow}
      \psi_{\alpha}^{\dagger}({\bf x}) \right.
      \left(- \frac{\hbar^2 \nabla^2}{2m} - \mu'_{\alpha} \right)
      \psi_{\alpha}({\bf x})                 \nonumber \\
 & &   +  \Delta_0 \psi_{\uparrow}^{\dagger}({\bf x})
              \psi_{\downarrow}^{\dagger}({\bf x})
   + \Delta_0^* \psi_{\downarrow}({\bf x})\psi_{\uparrow}({\bf x}) \nonumber \\
 & &   \left.     - \frac{|\Delta_0|^2}{V_0}
          - \frac{4\pi a\hbar^2}{m}~n_{\downarrow} n_{\uparrow} \right\}~,
\label{hamiltoniaan}
\end{eqnarray}
where $n_{\alpha} = \langle \psi_{\alpha}({\bf x}) \psi_{\alpha}({\bf x}) 
\rangle$\ is the equilibrium value of the
density of atoms in state $|\alpha\rangle$, and equivalently
$\Delta_0 = V_0 \langle \psi_{\downarrow}({\bf x})
\psi_{\uparrow}({\bf x}) \rangle$ is the equilibrium value of 
the BCS order parameter \cite{degennes}. The chemical
potential of each hyperfine state has now been renormalized to  
$\mu'_{\alpha} = \mu_{\alpha} - T^{2B} n_{-\alpha}$\ 
to include, on the mean-field level,
all two-body scattering processes with particles in state
$|-\alpha \rangle$. The factor $T^{2B} = 4\pi  a\hbar^2/m$\ is the
two-body scattering matrix, and has been substituted for $V_0$ to incorporate
correctly all two body processes into the calculation. Note that the
same substitution should
not be performed in the expression for $\Delta_0$, because all two-body
interactions are already going to be included by the BCS-treatment
as we will see below \cite{henk2}. 
Due to the nonequilibrium in the spin degrees of freedom, both 
chemical potentials $\mu'_{\downarrow}$
and $\mu'_{\uparrow}$\ need not be equal, and therefore the densities
of atoms in the respective hyperfine level can be varied independently.

Substituting for the operator $\psi_{\alpha}^{\dag}$\ the expression 
\begin{equation}
\psi_{\alpha}^{\dag} ({\bf x}) = \frac{1}{\sqrt{V}}\sum_{\bf k}
a_{{\bf k},\alpha}^{\dag} e^{-i {\bf k} \cdot {\bf x}},
\label{secondq}
\end{equation}
where $a^{\dag}_{{\bf k}, \alpha}$\ creates one particle in 
spin state $|\alpha \rangle$\ with momentum $\hbar {\bf k}$,  
the hamiltonian in Eq.~(\ref{hamiltoniaan}) becomes
\[
H  = \sum_{\bf k} \left( \begin{array}{cc}
a_{{\bf k},\uparrow}^{\dag}  & a_{-{\bf k},\downarrow} \end{array} \right)
\left( \begin{array}{cc} \epsilon_{\bf k} - \mu'_{\uparrow} & \Delta_0 \\
\Delta_0^* & -\epsilon_{\bf k} + \mu'_{\downarrow} \end{array} \right)
\left( \begin{array}{c} a_{{\bf k},\uparrow} \\ a_{-{\bf k},\downarrow}^{\dag}
\end{array} \right) 
\]
\begin{equation}
\hspace{3cm} -\frac{|\Delta_0|^2}{V_0} - n_{\uparrow}n_{\downarrow} T^{2B}
\label{nondiagH}
\end{equation}
where $\epsilon_{\bf k} = \hbar^2 {\bf k}^2/2m$\ is the free particle energy of
a particle with momentum $\hbar {\bf k}$.  
The density of atoms in state $| \alpha \rangle$\ is determined by
\begin{equation}
\label{n}
n_{\alpha} = \langle \psi_{\alpha}^{\dag}
                     \psi_{\alpha} \rangle =
                     \frac{1}{V}\sum_{\bf k} \langle
               a^{\dag}_{{\bf k},\alpha} a_{{\bf k},\alpha} \rangle.
\end{equation}
Since the effective mean-field hamiltonian in terms of the operators 
$a^{\dag}_{{\bf k},
\alpha}$\ and $a_{{\bf k},\alpha}$ is non-diagonal, one cannot directly
calculate the expectation value $\langle a^{\dag}_{{\bf k},\alpha} 
a_{{\bf k},\alpha} \rangle$.

This is, as usual, resolved by first applying a Bogoliubov transformation
according to \cite{degennes}
\begin{mathletters}
\begin{equation}
a_{{\bf k},\uparrow}  =  u_{\bf k} b_{{\bf k},\uparrow} + v^*_{\bf k}
                           b^{\dag}_{-{\bf k}, \downarrow} \label{Bogtrans1}
\end{equation}
\begin{equation}
a^{\dag}_{-{\bf k},\downarrow}  =  - v_{\bf k} b_{{\bf k},\uparrow} +  
                            u^*_{\bf k} b^{\dag}_{-{\bf k},\downarrow} 
\label{Bogtrans2}
\end{equation}
\label{Bogtrans}
\end{mathletters}

\noindent
to diagonalize the hamiltonian in Eq.~(\ref{nondiagH}). 
After performing this unitary transformation, we require 
that the hamiltonian in terms of the new quasiparticle operators 
$b_{{\bf k}, \uparrow}$\ and $b^{\dag}_{-{\bf k},\downarrow}$\ only has diagonal
elements, and furthermore that these operators again obey the usual 
anti-commutation relations for annihilation and creation operators. 
This determines the values of the yet
unknown, and in principle complex, constants $u_{\bf k}$\ and 
$v_{\bf k}$. The latter constraint requires that the constants 
$u_{\bf k}$\ and $v_{\bf k}$\ must satisfy the relations $|u_{\bf k}|^2 +
|v_{\bf k}|^2 =1$, and the requirement of diagonality of the
hamiltonian after the transformation leads to the condition
$|u_{\bf k}|^2 = \frac{1}{2}(1 + \xi_{\bf k}/
\sqrt{\xi^2_{\bf k}+ |\Delta_0|^2})$, introducing 
$\xi_{\bf k} = \epsilon_{\bf k} -\epsilon_F$, i.e.\ the free
particle energy relative to the average Fermi level $\epsilon_F = 
(\mu'_{\uparrow}+\mu'_{\downarrow})/2$. 

\begin{figure}[htbp]
\psfig{figure=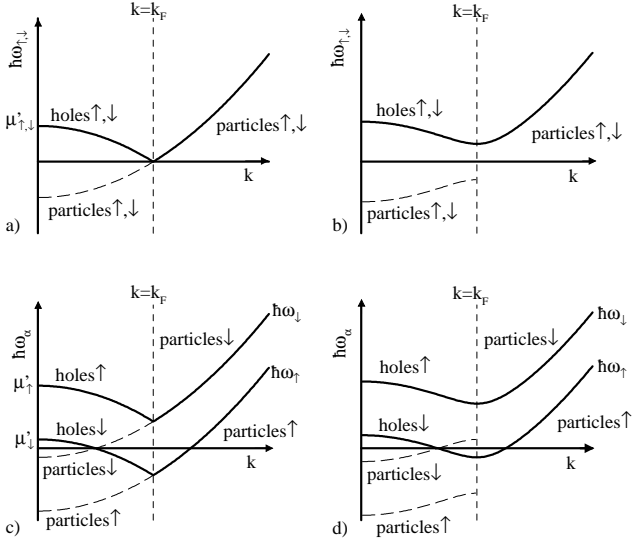}
\caption{\narrowtext
           Bogoliubov dispersion $\hbar \omega_{{\bf k}, \alpha}$\ for a)
$n_{\uparrow}=n_{\downarrow}$, $\Delta_0 =0$, b) $n_{\uparrow} = n_{\downarrow}$,
$\Delta_0 \neq 0$, c) $n_{\uparrow} > n_{\downarrow}$, $\Delta_0 = 0$, and
d) $n_{\uparrow} > n_{\downarrow}$, $\Delta_0 \neq 0$. The thin dashed lines
indicate the particle dispersions below the Fermi level $\epsilon_F$.}
\label{fig3}
\end{figure}

The eigenvalues corresponding to the Bogoliubov quasi particles are
then given by 
\begin{equation}
\hbar \omega_{{\bf k},\alpha} = -m_{\alpha} \delta \epsilon_F
+ \sqrt{\xi_{\bf k}^2 + |\Delta_0|^2}
\label{omega}
\end{equation}
where $m_{\alpha} = \pm 1/2$\ for $\alpha = \uparrow,\downarrow$,
respectively. Furthermore $\delta \epsilon_F = \mu'_{\uparrow} -
\mu'_{\downarrow}$\ is the difference in Fermi levels of the two hyperfine
states.
The dispersion relations of Eq.~(\ref{omega}) are depicted in Fig.~\ref{fig3}
for equal (Figs.~\ref{fig3}$a$ and \ref{fig3}$b$) and unequal 
densities (Figs.~\ref{fig3}$c$ and \ref{fig3}$d$) both
with zero (Figs.~\ref{fig3}$a$ and \ref{fig3}$c$) and nonzero 
$\Delta_0$\ (Figs.~\ref{fig3}$b$ and \ref{fig3}$d$),
respectively \cite{marianne}.

Note that, when the densities in both spin states are equal (corresponding
to $\delta \epsilon_F = 0$), the dispersion 
relation reduces to the usual Bogoliubov dispersion $\hbar \omega_{{\bf k}} =
\sqrt{\xi_{{\bf k}}^2 + |\Delta_0|^2}$\ 
describing particles above the Fermi level, i.e.\ $\epsilon_{\bf k} > 
\epsilon_F$, and holes (for which the
dispersion is given by minus the particle dispersion) below $\epsilon_F$.  
It is clear that the Bogoliubov transformation couples 
particles in state $|\alpha \rangle$\ with holes in state 
$|- \alpha \rangle$, see for example Fig.~\ref{fig3}$d$, and that for unequal
densities the dispersion relations are shifted with a constant
$\pm \delta \epsilon_F /2$ such that there 
appear two separate branches in the excitation spectrum of the Bogoliubov
quasi particles as shown in Figs.~\ref{fig3}$c$ and \ref{fig3}$d$.
For $n_{\uparrow} > n_{\downarrow}$, the negative sign
of $\hbar \omega_{{\bf k}, \uparrow}$\ around the Fermi level 
$\epsilon_F$\ indicates that the energy states are partially filled with spin 
down holes below $\epsilon_F$,
and with spin up electrons in a small region above the Fermi level.  
Therefore, the lower branch is {\it gapless} when $\Delta_0 < 
\delta \epsilon_F/2$, whereas the upper one always has a gap, even at 
$\Delta_0 = 0$. The case of unequal densities is thus analogous to a gapless
superconductor.

By plugging the transformation Eq.~(\ref{Bogtrans}) into Eq.~(\ref{n}),
it is easily verified that the densities satisfy
\begin{equation}
n_{\alpha} = \frac{1}{V} \sum_{\bf k} \left\{ |u_{\bf k}|^2 N(\hbar \omega_{{\bf k},
\alpha}) + |v_{\bf k}|^2 [1 - N(\hbar \omega_{{\bf k}, - \alpha})] \right\},
\label{n1}
\end{equation}
where $N(\hbar \omega_{{\bf k},\alpha}) = 
1/(\exp{[\beta \hbar \omega_{{\bf k},
\alpha}]} + 1) = \langle b^{\dag}_{{\bf k},\alpha} b_{{\bf k},\alpha} \rangle$\ 
is the Fermi distribution for the Bogoliubov quasi particles, and $\beta=1/k_BT$.
For fixed $n_{\alpha}$, Eq.~(\ref{n1}) determines the chemical potentials
$\mu'_{\alpha}$\ of the particles in state $|\alpha \rangle$.

Subsequently, the equilibrium value of the BCS order parameter is calculated
from $\Delta_0 = V_0 \langle \psi_{\downarrow}({\bf x}) \psi_{\uparrow} ({\bf x})
\rangle$.
Substituting Eqs.~(\ref{secondq}) and (\ref{Bogtrans}) for $\psi_{\uparrow,
\downarrow}({\bf x})$, this leads to the following BCS `gap equation'
\begin{equation}
\frac{1}{V} \sum_{\bf k}  \frac{1-N(\hbar 
\omega_{{\bf k},\uparrow}) -N(\hbar \omega_{{\bf k},\downarrow})}
{2\sqrt{\xi^2_{\bf k} + |\Delta_0|^2} }
 = - \frac{1}{V_0}.
\label{gapeq1}
\end{equation}
This equation has an ultra-violet divergence, as a consequence
of the fact that we made the assumption that the interparticle interaction
is local, i.e.~momentum independent. However, from the Lippmann-Schwinger
equation for the two-body scattering matrix \cite{glockle} 
\begin{equation}
\frac{1}{T^{2B}} = \frac{1}{V_0} + \frac{1}{V} \sum_{\bf k} 
\frac{1}{2 \xi_{\bf k}},
\label{LS}
\end{equation}
we find that this divergence is cancelled by a renormalization of $1/V_0$\
to $1/T^{2B}$\ \cite{henk2}, and the gap equation becomes
\begin{equation}
\frac{1}{V} \sum_{\bf k}  \left\{ \frac{1-N(\hbar 
\omega_{{\bf k},\uparrow}) -N(\hbar \omega_{{\bf k},\downarrow})}
{2\sqrt{\xi^2_{\bf k} + |\Delta_0|^2} } - \frac{1}{2\xi_{\bf k}} \right\}
 = - \frac{1}{T^{2B}}.
\label{gapeq}
\end{equation}
Eliminating from this equation both chemical potentials $\mu'_{\alpha}$\
by means of Eq.~(\ref{n1}), and equating 
$\Delta_0$\ to zero, one finds
the critical temperature $T_c$\ as a function of both hyperfine densities
in the gas. If the hyperfine densities are taken to be equal, the critical
temperature can be calculated analytically \cite{melo}, resulting in  
\begin{equation}
\label{tc}
T_c \simeq \frac{8\epsilon_F}{k_B \pi} e^{\gamma-2}\ 
             \exp \left\{ -\frac{\pi}{2k_F|a|}  \right\}~,
\end{equation}
where $\gamma=0.5772$\ is Euler's constant and 
$k_F= \sqrt{2 m \epsilon_F}/\hbar$ is again the wave vector corresponding to
the Fermi energy $\epsilon_F$. Including fluctuations 
changes only the prefactor of Eq.~(\ref{tc}) \cite{gorkov}. Although this is
expected to lower the
critical temperature somewhat, the exponential dependence of $T_c$
on the scattering length is most important for our purposes. 
Since taking fluctuations into account self-consistently is rather difficult,
in particular in the inhomogeneous case, 
we will come back to the effect of fluctuations
on the transition to a superfluid state in a future publication and consider
here only the mean-field theory, which is also known as the
many-body $T$-matrix theory. 

\begin{figure}[htbp]
\psfig{figure=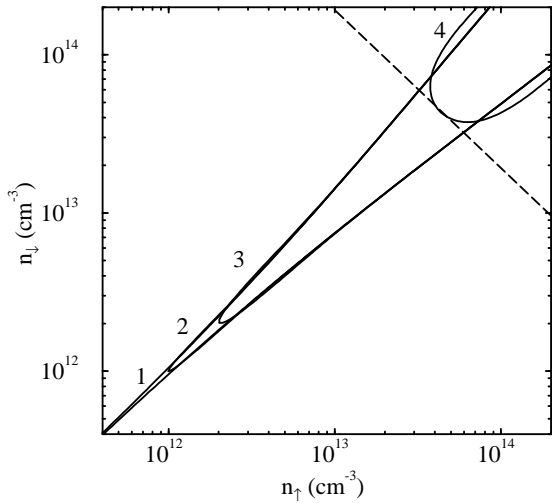}
\caption{\narrowtext
Contours of the critical temperature as a function of the
hyperfine densities $n_{\uparrow}$\ and $n_{\downarrow}$, for
1) $T=0.01$ nK, 2) $T=11$\ nK, 3) $T=37$\ nK, and 4) $T=1725$ nK. 
The dashed line is the spinodal line.}
\label{fig4}
\end{figure}

As mentioned previously, the densities of particles, and hence
the chemical potentials, need not be equal in both spinstates. 
In Fig.~\ref{fig4} we plot several contour plots of 
the critical temperature for the homogeneous gas in the $n_{\uparrow}-
n_{\downarrow}$\ plane. As can be seen from this figure, the most 
favorable situation is that, given a certain total density of atoms, 
both hyperfine states are equally occupied because this gives
rise to the highest critical temperature.   
When the two hyperfine states are not equally occupied, it can be shown
that there is a nonzero critical temperature only when the
spin `polarization' $|n_{\uparrow}-n_{\downarrow}|/(n_{\uparrow} +
n_{\downarrow}) < 3 k_BT_c/2\epsilon_F$. Also, for fixed average 
Fermi level $\epsilon_F$ and increasing difference
$\delta \epsilon_F$, the critical temperature decreases, and there 
is no transition at all when $\delta \epsilon_F \stackrel{>}{\sim}
k_B T_c(0)$,
with $T_c(0)$\ the critical temperature when $\delta \epsilon_F=0$\ \cite{henk}. 
This behavior is similar to what occurs in superconductors placed in a magnetic field
and can be understood physically from the fact that the formation of Cooper pairs 
spreads the occupation of energy levels 
only over an energy interval of order $\Delta_0 \simeq k_BT_c$ around the
respective Fermi levels $\mu'_{\uparrow}$\ and $\mu'_{\downarrow}$. 
Moreover, pairing between atoms at the average Fermi energy can only
take place if there exists an overlap
between the Fermi distributions of the two spin states in this region of
momentum space. This
indeed shows that $\delta \epsilon_F$\ must be smaller than about $k_BT_c(0)$.  

The dashed line in Fig.~\ref{fig4}\ is the spinodal line, above which
the gas becomes mechanically unstable. We will return to this issue in 
the next section.

\subsection{Mechanical stability of a two-component Fermi gas}
\label{stability}

As was already pointed out in Ref.~\cite{henk}, an important requirement for a BCS
transition to occur, is that the system is mechanically stable against
density fluctuations. The negative $s$-wave scattering length induces an
effectively attractive interatomic potential, so if the density of particles
becomes too large, the system can collapse to a fluid or solid state before
the systems becomes superfluid in the (metastable) gaseous phase. 
In general, for mechanical stability of the gas at the critical temperature, 
we must require
that the velocities of the two sound modes in the normal state of the
gas are real. These
velocities can be calculated from the free energy density $f$ of the gas. Since
the temperatures of interest are so low that $k_BT\ll \epsilon_F$, 
we can consider the zero-temperature
limit, in which the free energy density amounts to the average energy 
density $f= \langle E \rangle/V$. We thus have 
\begin{eqnarray}
f& = & f_0 + f_{int}
\nonumber \\
& \equiv & \frac{3}{10} (6\pi^2)^{2/3} (n_{\uparrow}^{5/3} + n_{\downarrow}^{5/3})
\frac{\hbar^2}{m} + n_{\uparrow} n_{\downarrow} T^{2B},
\label{vrijeenergie}
\end{eqnarray}
where $f_0$\ is the ideal gas free energy density of the particles in each
hyperfine state at $T=0$, and $f_{int}$\ is the free energy density that 
arises due to interactions between particles in both spin states.
The corresponding sound velocities squared are determined by the eigenvalues
of the matrix 
\[
\left( \begin{array}{cc}
        \frac{\partial^2 f}{\partial n_{\uparrow}\partial n_{\uparrow}} &
        \frac{\partial^2 f}{\partial n_{\uparrow}\partial n_{\downarrow}} \\ 
        \frac{\partial^2 f}{\partial n_{\downarrow}\partial n_{\uparrow}} &
        \frac{\partial^2 f}{\partial n_{\downarrow}\partial n_{\downarrow} } 
       \end{array} \right) ,
\]
leading to the condition that $n_{\uparrow} n_{\downarrow} a^6 \leq (\pi/48)^2$.
The line in the $n_{\uparrow}-n_{\downarrow}$ plane, where the equality
holds, is called the spinodal line, and for the homogeneous $^6$Li gas it
is plotted as the dashed line in Fig.~\ref{fig4}.

Notice, however, that a spin-polarized Fermi gas becomes unstable at
densities above the spinodal line, irrespective of the sign of the
scattering length $a$. Therefore the question arises as to
what exactly happens
at densities above the spinodal line, and whether there is a difference
in the behavior for positive or negative $s$-wave scattering length.
First of all, notice that 
the matrix $\partial^2 f/\partial n_{\alpha} \partial n_{\beta}$\
has an eigenvalue $\lambda =0 $ at the spinodal point. The 
corresponding eigenvector ${\bf \hat{e}}_0$ points into the unstable direction of
the phase space. For equal densities of the two hyperfine states, 
it is straightforward to calculate that 
${\bf \hat{e}}_0 = 1/\sqrt{2}( \mp1,1)$, where the upper and lower sign
refer to positive and negative scattering length $a$, respectively. 
We therefore conclude that for a negative $s$-wave scattering length,
the gas collapses to a dense phase (probably a solid), whereas
for positive $a$ it phase separates into two dilute gaseous phases with 
opposite `magnetization'. Since the last situation might be of interest
for future experiments with other fermionic atoms than $^6$Li, 
we consider now for a moment also the $a>0$ case. 

\subsubsection{The $a>0$ case}

To analyze the stability at positive $a$, we notice that the pressure 
of the gas at zero 
temperature is given by $p = -\partial \langle E \rangle/\partial V$. 
We thus find that 
\begin{eqnarray}
p & = & p_0 + p_{int} \nonumber \\
& = & \frac{1}{5} (6\pi^2)^{2/3}
(n_{\uparrow}^{5/3} + n_{\downarrow}^{5/3}) \frac{\hbar^2}{m} + 
n_{\uparrow} n_{\downarrow} T^{2B}.
\label{druk} 
\end{eqnarray}
Introducing for future convenience dimensionless variables 
according to $x\equiv n_{\uparrow} a^3$, $y \equiv n_{\downarrow}a^3$, 
$M_{\uparrow,\downarrow} \equiv (2m a^2/\hbar^2)
\mu_{\uparrow,\downarrow}$, $P\equiv a^3 (2ma^2/\hbar^2)p$,
and $F \equiv a^3(2ma^2/\hbar^2)f$, it follows 
from Eqs.~(\ref{druk}) and (\ref{vrijeenergie}) that 
\begin{eqnarray}
\label{drukxy}
P(x,y) & = & \frac{2}{5} (6\pi^2)^{2/3} ( x^{5/3} + y^{5/3}) + 8 \pi xy \\
\label{Fxy}
F(x,y) & = & \frac{3}{5} (6\pi^2)^{2/3} ( x^{5/3} + y^{5/3}) + 8 \pi xy \\
\label{muupxy} 
M_{\uparrow}(x,y) & = & (6\pi^2)^{2/3} x^{2/3} + 8\pi y \\
\label{mudownxy}
M_{\downarrow}(x,y) & = & (6\pi^2)^{2/3} y^{2/3} + 8\pi x, 
\end{eqnarray}
where we used that $\mu_{\uparrow,\downarrow} = \partial f/\partial
n_{\uparrow,\downarrow}$. Notice that these equations are symmetric under the 
exchange of the variables $x$\ and $y$, or rather the index $\uparrow$\ and 
$\downarrow$. 

\begin{figure}[htbp]
\psfig{figure=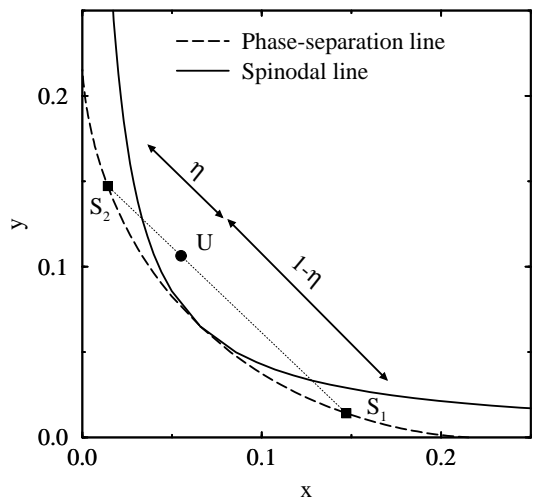}
\caption{\narrowtext
Plot of phase-separation line (dashed) as a function of the dimensionless
densities $x=n_{\uparrow}a^3$\ and $y=n_{\downarrow} a^3$, together with
the spinodal line $xy=(\pi/48)^2$, above which the gas phase separates
to the dashed line. As an example, the unstable phase $U$ separates
to the stable phases $S_1$\ and $S_2$, with volume fractions
$V_1=\eta V$\ and $V_2 = (1-\eta)V$, respectively.
Note that in the regions between the phase-separation line
and the spinodal line, the gas is metastable.}
\label{fig5}
\end{figure}

The condition that must be fulfilled for a phase separation, is that
an unstable phase $U$\ separates into two distinct phases $S_1$\
and $S_2$\ in the stable region of the phase space, in such a way that both the
pressure as well as the chemical potential in the two stable phases are equal.
Since in our case we are dealing with a gas consisting of two constituents,
we require that both chemical potentials $\mu_{\uparrow}$\ and
$\mu_{\downarrow}$ must be equal in the two stable phases, otherwise 
particles would still
prefer one phase above the other, and there would be no equilibrium.  
A third condition that must hold is that the total number of particles in
each spin state must be conserved. In Fig.~\ref{fig5} we show the 
spinodal line, in terms of the dimensionless variables $x,y$, 
i.e.~$xy=(\pi/48)^2$. 
Furthermore we plotted an unstable point $U$, which separates into 
points $S_1=(x_1,y_1)$\ and $S_2= (x_2,y_2)$\ in the stable 
regime of phase space.
Next we will deduce the exact position of these points $S_1$\ and $S_2$\ 
from the above mentioned conditions on the phase separation.   

From the condition on the pressure and the symmetry of Eq.~(\ref{drukxy}) it
follows that $P_{S_1} = P(x_1,y_1) =P_{S_2}=P(x_2,y_2)$ is satisfied
if $x_1=y_2$\ and $x_2= y_1$. In other words, the separation points 
$S_1$\ and $S_2$\ lie symmetric in the $n_{\uparrow}-n_{\downarrow}$ plane.
The condition on the chemical potentials, i.e.~$M_{\uparrow,
\downarrow}(x_1,y_1) =M_{\uparrow,\downarrow}(x_2,y_2)$   
now determines the exact position of the points $S_1=(x_1,y_1)$ and 
$S_2=(x_2,y_2)$. From the symmetry of 
Eqs.~(\ref{muupxy}) and (\ref{mudownxy}) we see that 
$M_{\uparrow}(x_1,y_1)  
= M_{\downarrow}(y_1,x_1) = M_{\downarrow}(x_2,y_2)$,
so $M_{\uparrow}(x_i,y_i)=M_{\downarrow}(x_i,y_i)$\ in each
individual separation point $S_i,\ i=1,2$. This shows that, in practice
we are looking for intersections of the curves $M_{\uparrow}(x,y)
= M_{\downarrow}(x,y) = M$. Again from symmetry, it 
is immediately clear that
there is always a point of intersection of the two curves somewhere on the 
line $x=y$, but for certain values of $M$ there can be 
two additional points of intersection, which are plotted as the dashed line
in Fig.~\ref{fig5}. This line is the phase-separation line. As we will prove
later on, it coincides with the spinodal line at $x=y=\pi/48$, and lies below
the spinodal line, in the stable region of phase space elsewhere. 

The third condition requiring conservation of the total number of particles
in each spin state determines the volume fractions $V_1/V$\ and 
$V_2/V$\ of the two phases. For an unstable homogeneous system of volume 
$V$, and with $N_{\uparrow} = n_{\uparrow}^{U} V$\ and $N_{\downarrow} = 
n_{\downarrow}^{U} V$\ particles in the two hyperfine states, we have
that after the phase separation  
\begin{eqnarray}
N_{\uparrow} & = & n_{\uparrow}^{S_1} V_1 + n_{\uparrow}^{S_2} V_2 \nonumber \\
N_{\downarrow} & = & n_{\downarrow}^{S_1} V_1 + n_{\downarrow}^{S_2} V_2. 
\nonumber
\end{eqnarray}
Of course, the total density is also constant so we have 
$n_{total} = n_{\uparrow}^U + n_{\downarrow}^U = n_{\uparrow}^{S_1} + 
n_{\downarrow}^{S_1} = n_{\uparrow}^{S_2} + n_{\downarrow}^{S_2}$, 
which means that the points $U,\ S_1$\ and $S_2$ must lie on a straight
line given by $n_{\uparrow} + n_{\downarrow} = n_{total}$,
as is indicated for the points $U$, $S_1$, and $S_2$\ by the dotted line in
Fig.~\ref{fig5}. Defining now $\beta^{U} = 
n_{\uparrow}^{U}/n_{total}$, $\beta^{S} = n_{\downarrow}^{S_1}/n_{total} =
n_{\uparrow}^{S_2}/n_{total}$, and $\eta = (\beta^{U} - \beta^{S})/
(1 - 2 \beta^{S})$, we find after a little algebra that $V_1 = \eta V$,
and $V_2=(1-\eta) V$. So, the phase separation is such that for
arbitrary position of the point $U$ on the unstable
part of the dotted line in Fig.~\ref{fig5}, the system separates to the
same two stable points $S_1$\ and $S_2$; the exact position of $U$ determines
only the volume fractions of the stable phases. The phase points $S_1$\
and $S_2$\ have the same total density but differ in `spin magnetization'
by an amount
$| n_{\uparrow}^{S_1} - n_{\downarrow}^{S_1} |$. Therefore, 
the phase separation corresponds to a spin decomposition that is driven by the
fact that at sufficiently high densities the loss in interaction energy between
the two species compensates for the gain in kinetic energy due to the
Pauli exclusion principle.    
 
\begin{figure}[htbp]
\psfig{figure=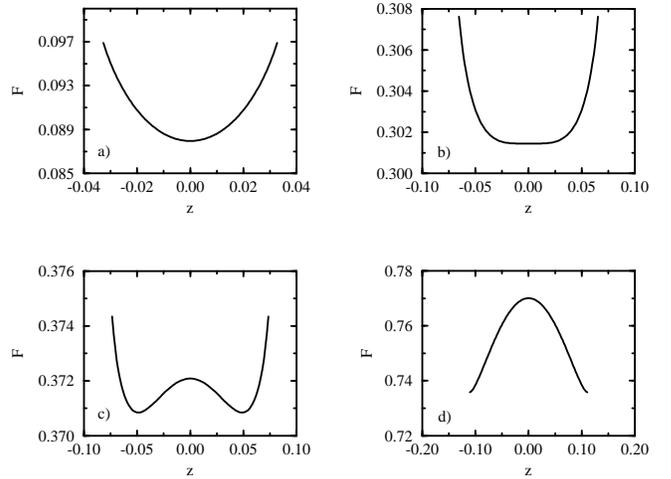}
\caption{\narrowtext
Plots of the dimensionless free energy $F(n,z)$ as a function of $z=
(y-x)/2$, for $a)$\ $n=\pi/96 < n_{sp}$, $b)\ n=\pi/48=n_{sp}$,
$c)\ n_{sp} < n=6\pi/256 < n_c$, and $d)\ n=9\pi/256=n_c$.}
\label{fig6}
\end{figure}

To gain even more understanding in this phase separation, and to distinguish
later on the situation with negative $a$ from the case with positive $a$,
we consider the dimensionless free energy in Eq.~(\ref{Fxy}) more closely.
It is clear from Fig.~\ref{fig5}\ that the phase separation takes
place on lines $x+y=\mbox{constant}$. Therefore we introduce new variables
$n$\ and $z$\ such that
\begin{eqnarray}
x& = & n-z \nonumber \\
y& = & n+z, \nonumber
\end{eqnarray} 
i.e.~the $n$-axis lies along the line $x=y$\ in Fig.~\ref{fig5}, and the 
$z$-axis lies along the line $y=-x$. Lines of constant $x+y$\ therefore
have a constant $n$ (density), and run parallel to the $z$-axis.
The dimensionless free energy $F(x,y)$\ in terms of these new variables
now becomes
\begin{eqnarray}
F(n,z) & = & \frac{3}{5} (6\pi^2)^{3/2} \left[ (n-z)^{5/3} + (n+z)^{5/3}
\right] \nonumber \\ 
& & \hspace{1in}  + 8\pi (n^2 -z^2).
\label{Fzn}
\end{eqnarray}
Note that, since the original variables $x$\ and $y$\ must be positive,
also $n \geq 0$, and for given $n$, we have $-n \leq z \leq +n$. 
Taking the derivative of $F(n,z)$\ with respect to $z$\ at constant $n$,
it is found that 
\[
\frac{\partial F}{\partial z} = (6\pi^2)^{2/3} \left[-(n-z)^{2/3} +
(n+z)^{2/3} \right] -16 \pi z , 
\]
which is zero
at $z=0$\ for all values of $n$. Hence, there is always an extremum
in the free energy $F(n,z)$ at the line $z=0$. To see whether this is a
minimum or a maximum, we have to analyze the second derivative
\[
\left. \frac{\partial^2 F}{\partial z^2} \right|_{z=0} = 
\frac{2}{3} (6\pi^2)^{2/3} \left[ \frac{2}{n^{1/3}} \right] - 16\pi,
\]
which is positive for $n<n_{sp} = \pi/48$, zero at $n=n_{sp}$, 
and negative for $n>n_{sp}$. So, the minimum in the free energy $F(n,z)$\ 
at constant $n$\ and
$z=0$\ changes into a maximum at $n=n_{sp}$, which exactly coincides with
the spinodal point at $x=y$. This behavior is shown in Fig.~\ref{fig6},
where we plot $F(n,z)$\ for $a)\ n<n_{sp},\ b)\ n=n_{sp},\ c)$\ and $d)\ 
n>n_{sp}$, as a function of $z$.  

From Fig.~\ref{fig6} we see that the maximum at $z=0$ for fixed
$n>n_{sp}$\ is flanked by two minima in the free energy, which move outward
in the $\pm z$-direction for increasing $n$. Moreover, for $n=n_c = 9\pi/256$\
the minima just appear at $z=\pm n$, i.e.\ at the $y$-axis in $y= 9\pi/128$ 
and at the $x$-axis in $x=9\pi/128$, respectively,
in the original dimensionless density variables $x$\ and $y$. The important 
point is now that these two minima in the free energy $F(n,z)$\ for fixed $n$
are, after transforming back to $x-y$ coordinates, precisely the stable separation 
points $S_1$\ and $S_2$. Because of symmetry, they obey all conditions 
that we imposed on them. Furthermore we notice that for $n>n_c$, or total
density $n_{total} \geq 9\pi/ 128 a^3$, the spin 
separation is complete, i.e.~one part of the volume is occupied with
only atoms in hyperfine level $|\uparrow \rangle$, the rest of the volume
only contains atoms in state $|\downarrow \rangle$. The densities of both
phases is in this case evidently $n_{\uparrow}^{S_1} = n_{\downarrow}^{S_2}
= n_{total}$.   

\subsubsection{The $a<0$ case}
We now consider the case where the scattering length $a<0$,
as is the case for the $^6$Li system. Introducing again dimensionless
variables according to $x=n_{\uparrow}|a|^3$\ and $y=n_{\downarrow}|a|^3$,
and after the substitution $x=n-z$\ and $y=n+z$, respectively, the dimensionless
free energy is readily seen to be 
\begin{eqnarray}
F(n,z) & = & \frac{3}{5} (6\pi^2)^{3/2} \left[ (n-z)^{5/3} + (n+z)^{5/3}
\right]  \nonumber \\ 
& & \hspace{1in}  - 8\pi (n^2 -z^2).
\label{Fznnegative}
\end{eqnarray}
The first derivative of $F$\ in the $z$-direction is given by
\[
\frac{\partial F}{\partial z} = (6 \pi^2)^{3/2} \left[ -(n-z)^{2/3}
+ (n+z)^{2/3} \right] + 16\pi z,
\]
which is always zero at $z=0$. The second derivative with respect to the variable
$z$ at $z=0$ is given by 
\[
\frac{\partial^2 F}{\partial z^2} = \frac{2}{3} (6 \pi^2)^{2/3}
\left[ \frac{2}{n^{1/3}} \right] + 16 \pi,
\]
which is for all allowed values of $n$ larger than zero. Therefore we conclude 
that there indeed can be no phase separation in the $z$-direction along the lines 
$n=\mbox{constant}$\ as was the case for positive $a$. 
 
Instead, the phase separation in the unstable region of phase space above
the spinodal line takes place in the $n$-direction. This can be shown by
considering the second derivative of $F$\ with respect to $n$, i.e.
\[
\frac{\partial^2 F}{\partial n^2} = \frac{2}{3} (6 \pi^2)^{2/3} \left[
\frac{1}{(n-z)^{1/3}} + \frac{1}{(n+z)^{1/3}} \right] -16 \pi  
\]
which at $z=0$, or $x=y$, becomes zero exactly at $n=n_{sp}= \pi/48$. The
fact that the second derivative of the free energy is zero at some point
signals an instability in that direction (in the $a>0$ case, the
second derivative of $F$\ with respect to $z$ just became zero at $n=n_{sp}$).
So we find that in the case of negative scattering length, the unstable
point $U$ in phase space will separate into a phase $S_1$ with lower total
particle density, and a phase $S_2$ with higher total particle density, or
in other words, to a gaseous and a dense (solid) state. However, we do not
have an appropriate theory that can also describe the dense phase. 
Therefore we do not consider this
kind of phase separation, which is very common in gases and liquids, 
further here. 
 
\section{INHOMOGENEOUS FERMI GAS}
\label{inhomogeen}

\subsection{Local density approximation}
\label{LDA}
Until now we only considered a homogeneous gas of spin-polarized
atomic $^6$Li. In reality, however, experiments with ultra-cold atomic gases 
are performed by trapping and evaporatively cooling the gas in an external 
potential which generally can be modeled
by an isotropic harmonic oscillator $V({\bf r}) =
\frac{1}{2} m \omega^2 {\bf r}^2$, where $\omega$\ is the trapping frequency. An
exact calculation of the (inhomogeneous) density of the gas by calculating
all eigenstates of the trapping potential is very elaborate but
has nevertheless been performed for the bosonic isotopes $^7$Li \cite{mh,bergeman}
and $^{87}$Rb \cite{hutchinson}. Fortunately,
in the fermionic system it is a good approximation to make use of the
local density approximation, which treats the system as being locally 
homogeneous. This requires in the first place that the correlation length $\xi
= {\cal O}(1/k_F)$\ is much shorter than the length scale 
$l= \sqrt{\hbar/m \omega}$\ over which the densities change. This condition
is equal to the condition that the level spacing $\hbar \omega$ of the
trapping potential is much smaller than the Fermi energy.  
Secondly, below the critical temperature, the size of the Cooper pairs
must be smaller than $l$, or else the trapping potential would
influence the wave function of the Cooper pairs. This size is essentially 
temperature-independent and of the order of ${\cal O}(\hbar v_F/\pi 
\Delta_0(0))$, where $\Delta_0(0)$\ is the zero-temperature value
of the BCS order parameter, and $v_F = \hbar k_F/m$\ the Fermi 
velocity corresponding to $\epsilon_F$.  
Of course, the local density approximation always breaks down
at the edge of the gas cloud where the density vanishes and
the effective Fermi energy becomes zero, and also
at the critical temperature where the correlation length $\xi$ diverges. 
So at a nonzero temperature below $T_c$ there are two spatial regions 
where the local density approximation is not valid, 
i.e.~around the position where the local BCS order parameter vanishes 
and around the position where the local Fermi energy vanishes.
However, these regions are 
so small, that we do not expect any important changes in 
the functional behavior of physical properties at the crossover from 
outside to inside these regions. As a result we believe that it is 
rather accurate to apply the local density approximation 
to calculate $T_c$ \cite{giorgino,mh2}.

In this approximation, the densities 
$n_{\uparrow}$\ and $n_{\downarrow}$\ of the two hyperfine states
together with the gap $\Delta_0$\ can still be calculated by means of
the equations derived in Sec.~\ref{BCStransition}, with the understanding that
now the effective chemical potentials, and consequently the densities 
and $\Delta_0$, are spatially dependent through 
\begin{equation}
\mu'_{\alpha}({\bf r}) = \mu_{\alpha} - V({\bf r}) -
n_{-\alpha}({\bf r}) T^{2B},
\label{mueffectief}
\end{equation}
where $\mu_{\alpha}$\ is the overall (constant) bare chemical potential of
atoms in hyperfine state $|\alpha \rangle$. So, given $T$, $\mu_{\downarrow}$\ 
and $\mu_{\uparrow}$\ (or equivalently $T$, $N_{\downarrow} = \int d{\bf r} \
n_{\downarrow}({\bf r})$\ and $N_{\uparrow} = \int d{\bf r}\ 
n_{\uparrow}({\bf r})$), one can 
determine the values of $n_{\downarrow}({\bf r})$, $n_{\uparrow}({\bf r})$, 
and $\Delta_0({\bf r})$\ self-consistently for every position 
${\bf r}$\ in space, as if the system were homogeneous.
This procedure will be used in the next section to calculate 
the critical temperature of the spin-polarized gas as a function of the number
of particles in the trap. 

\subsection{Critical temperature}
\label{Tc}

\begin{figure}[htbp]
\psfig{figure=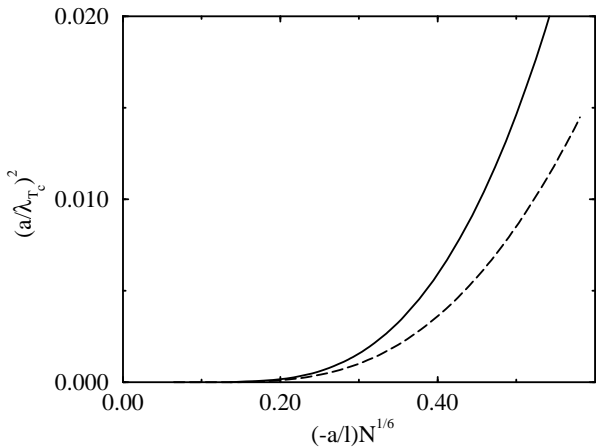}
\caption{\narrowtext
Critical temperature as a function of the number of particles 
(solid line), when
there are $N$ particles present in both spin states. The dashed line represents 
the critical temperature for a gas whose density
distribution is not altered by mean-field interactions.}
\label{fig7}
\end{figure}

The critical temperature $T_c$\ of the gas is such that at the center
of the magnetic trap, where the density of the gas is highest, the 
energy gap $\Delta_0({\bf 0})$\ just becomes nonzero for a given 
number of particles $N_{\uparrow}$\ and $N_{\downarrow}$.   
First we will consider the case where $N_{\uparrow}=N_{\downarrow}=N$. In 
Fig.~\ref{fig7} the solid line shows the result of our
calculation. The dashed line in this figure gives
the critical temperature for the Fermi gas if one 
does not include the effects of the mean-field interaction in 
Eq.~(\ref{mueffectief}). In this approximation, 
the number of particles in each hyperfine state is, with a high degree of
accuracy, given by the zero-temperature result
$N_{\uparrow,\downarrow}=(\mu_{\uparrow,\downarrow}/\hbar \omega)^3/6$,
and the density in the center of the trap is $n_{\uparrow,\downarrow}({\bf 0})
= (2 m \mu_{\uparrow,\downarrow}/\hbar^2)^{3/2}/(6\pi^2)$,
which is considerably smaller than
in case that the mean-field interaction is taken into account. As a
result, the critical temperature obtained in this manner
is substantially lower for an equal number of particles. From an
experimental point of view, it is therefore important to include interactions 
to obtain a reliable estimate for the critical temperature as a function 
of the number of trapped particles.

We found that, as is also the case for a Bose gas in a harmonic trap 
\cite{giorgino}, the critical temperature, or rather the dimensionless parameter
$a/\lambda_{T_c}$\ is a universal function of $N^{1/6}a/l$, with
the thermal DeBroglie wavelength $\lambda_T=\sqrt{2\pi \hbar^2/ mk_BT}$\ and
$l=\sqrt{\hbar/m\omega}$. The solid line in Fig.~\ref{fig7} is can be fitted
numerically very well with the expression
\[
\left( \frac{a}{\lambda_{T_c}}\right)^2 = 
0.037 \exp{\left\{ -1.214 \frac{l}{|a|}N^{-\frac{1}{6}} +
2.990 \frac{|a|}{l} N^{\frac{1}{6}} \right\} }, 
\]
for the whole range of parameters shown in Fig.~\ref{fig7}.

The fact that the critical temperature is a universal
function of the parameter $N^{1/6}a/l$ can be understood easily by rewriting 
the gap equation Eq.~(\ref{gapeq}) at the critical temperature
in the form
\[
\frac{\sqrt{\pi}}{4} \frac{\lambda_{T_c}}{a}   =  
\hspace{2in}   
\]
\[
\int_0^{\infty} dx \sqrt{x} 
\frac{ N(- \delta \epsilon_F + |\frac{x}{\beta}-\epsilon_F|) + 
N(\delta\epsilon_F + |\frac{x}{\beta}-\epsilon_F|)}{2|x-\beta \epsilon_F |}, 
\]
where $N(x) = 1/(\exp{[\beta x]}+1)$ is the Fermi distribution.
This shows that at the critical temperature, 
$a/\lambda_{T_c}$\ is
a function of $\delta\epsilon_F/k_BT_c$\ and $\epsilon_F/k_B T_c$\ only.
Equivalently, from the density for each spinstate given in Eq.~(\ref{n1}), 
and the fact that at the critical temperature the densities in the center
of the trap $n_{\alpha}({\bf 0})$ are critical, we find that
\[
n_{\alpha} ({\bf 0}) \lambda_{T_c}^3 = F_\alpha \left[ \frac{\delta \epsilon_F}{
k_BT_c}, \frac{\epsilon_F}{k_BT_c} \right].
\] 
So the central density of each spin state times the thermal wavelength
is also a function of the dimensionless parameters $\delta\epsilon_F/
k_BT_c$\ and $\epsilon_F/k_BT_c$. Combining these two equations, it
follows that $a/\lambda_{T_c}$\ is directly related to the densities
in the center of the trap, i.e.
\begin{equation}
\frac{a}{\lambda_{T_c}} = F [ n_{\uparrow}({\bf 0}) \lambda_{T_c}^3,  
n_{\downarrow}({\bf 0}) \lambda_{T_c}^3].
\label{aoverlambdaTc}
\end{equation}

To prove now that $a/\lambda_{T_c}$\ is a function of $N^{1/6} a/l$, it should
be noticed that in general in the local density approximation for $T\geq T_c$
\begin{equation}
n_{\alpha}({\bf r})\lambda_{T}^3 = f_{3/2} \left[ \exp{\{ \beta [
\mu_{\alpha} - n_{-\alpha}({\bf r}) T^{2B} - V({\bf r}) ] \} } \right],
\label{fermifunctie}
\end{equation}
where $f_{3/2}[z({\bf r})]$ is the Fermi function originating from 
integration over momenta and analogous to the Bose function $g_{3/2}(z)$. 
Applying this equation at ${\bf r}={\bf 0}$\ and $T=T_c$,
we find that both chemical potentials are functions of $a/\lambda_{T_c}$
and the central densities of both hyperfine states, and obey
\[
\frac{\mu_{\alpha}}{k_BT_c} = F'_{\alpha} \left[ \frac{a}{\lambda_{T_c}},
n_{\uparrow}({\bf 0}) 
\lambda_{T_c}^3, n_{\downarrow}({\bf 0})\lambda_{T_c}^3 \right].
\]
For a general value of ${\bf r}$, but still at $T=T_c$, we can apply
the substitution 
\begin{equation} 
y = \sqrt{\frac{m \omega^2}{2 k_B T_c} } r 
\label{substitutie}
\end{equation}
in Eq.~(\ref{fermifunctie}), from which it follows immediately that for 
each hyperfine state 
\[
n_{\alpha} ({\bf r}) \lambda_{T_c}^3 = F_{\alpha}'' \left[ \frac{a}{\lambda_{T_c}},
n_{\uparrow}({\bf 0}) \lambda_{T_c}^3, n_{\downarrow}({\bf 0}) 
\lambda_{T_c}^3, y^2 \right].
\]
To find the total number of particles in each hyperfine level, we then 
integrate this result over the spatial extent of the gas cloud, resulting in
\begin{eqnarray}
N_{\alpha} & = & 4 \pi \int_{0}^{\infty} dr \ r^2 n_{\alpha}(r) \nonumber \\
& = & \frac{4 \pi}{\lambda_{T_c}^3} \left( \frac{2 k_B T_c}{
m \omega^2} \right)^{3/2} \left. \int dy \ \right\{ y^2 \times \nonumber \\
& & \ \ \ \ \ \ \ \ \ \ \ \ \ \ \ \left.  F_{\alpha}''
\left[ \frac{a}{\lambda_{T_c}},
n_{\uparrow}({\bf 0}) \lambda_{T_c}^3, n_{\downarrow}({\bf 0}) 
\lambda_{T_c}^3, y^2 \right] 
\right\} \nonumber \\
& \equiv & \left( \frac{l}{\lambda_{T_c}} \right)^6 F_{\alpha}''' 
\left[ \frac{a}{\lambda_{T_c}}, n_{\uparrow}({\bf 0}) \lambda_{T_c}^3, 
n_{\downarrow}({\bf 0}) \lambda_{T_c}^3 \right].
\label{Nupdown}
\end{eqnarray} 
Multiplying Eq.~(\ref{Nupdown}) on both sides with $(a/l)^6$, 
and using the result of Eq.~(\ref{aoverlambdaTc}), it is proven 
that at the critical temperature 
\begin{equation} 
\frac{a}{\lambda_{T_c}} = F \left[ N_{\uparrow}^{1/6} \frac{a}{l},
N_{\downarrow}^{1/6} \frac{a}{l} \right],
\label{universeel}
\end{equation}
so that, when $\mu_{\uparrow} = \mu_{\downarrow}$ the dimensionless
parameter $a/\lambda_{T_c}$\ is a universal function of $N^{1/6}a/l$. 
   
The spinodal point in this case is given by $N^{1/6} a/l \simeq 
0.66$, and is not included in Fig.~\ref{fig7}, because for $^6$Li
trapped in a harmonic potential with frequency $\nu = \omega/2\pi = 144$Hz,
or $\hbar \omega/k_B \simeq 6.9$ nK, 
corresponding to the present experimental conditions at the Rice experiment
\cite{rice}, 
spinodal decomposition only occurs with as many as $5.8 \times 
10^7$ particles.  

\begin{figure}[htbp]
\psfig{figure=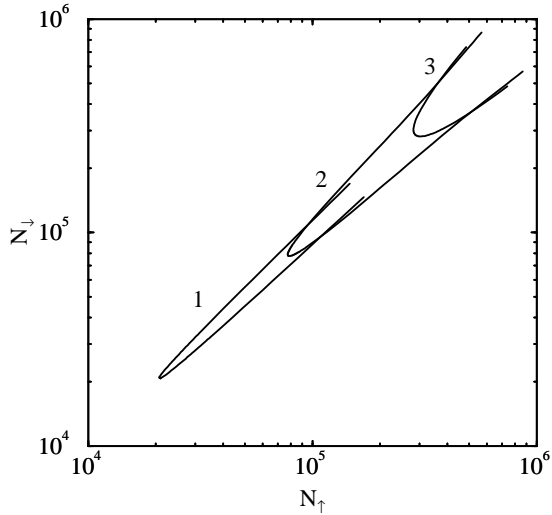}
\caption{\narrowtext
Critical temperature as a function of the number of $^6$Li atoms in each
hyperfine state. Curve 1 to 3 give the combinations 
$(N_{\uparrow},N_{\downarrow})$ corresponding to $T_c=3$~nK, 2)
$T_c=11$~nK, and 3) $T_c=37$~nK. For equal number of particles in
each hyperfine state, the density of particles in the center
of the trap corresponds to 1) $n_{\uparrow,\downarrow}({\bf 0})=
0.5\times 10^{12}$~cm$^{-3}$, 2) $1\times 10^{12}$~cm$^{-3}$, and 3)
$2\times 10^{12}$~cm$^{-3}$\ respectively.} 
\label{fig8}
\end{figure}

For an unequal number of particles in each hyperfine state, we 
find a universal surface for $a/\lambda_{T_c}$\ as a function
of $N_{\downarrow}^{1/6} a/l$\ and $N_{\uparrow}^{1/6}a/l$, as
Eq.~(\ref{universeel}) shows. However, since
we are in this paper mainly interested in trapping $^6$Li atoms, we will
calculate several contours of the critical temperature of such a gas
trapped in an isotropic harmonic oscillator with $\nu =144$Hz.
The results are plotted in Fig~\ref{fig8}.  
Again we see that given the total number of particles in the gas,
the most favorable situation is the one with equal numbers of particles
in each hyperfine state.

An important experimental question is how we could observe
whether or not the gas is superfluid at a certain temperature. 
An immediate possibility that, in view of the results with the BEC experiments,
comes to mind is to consider whether there is a change in the
density profile at the critical temperature. In the next section we
will therefore concentrate on the superfluid state of the gas, and
determine the density profiles and in addition
the spatial dependence of the energy gap $\Delta_0({\bf r})$. 

\subsection{Superfluid state}
\label{benedenTc}

\begin{figure}[htbp]
\psfig{figure=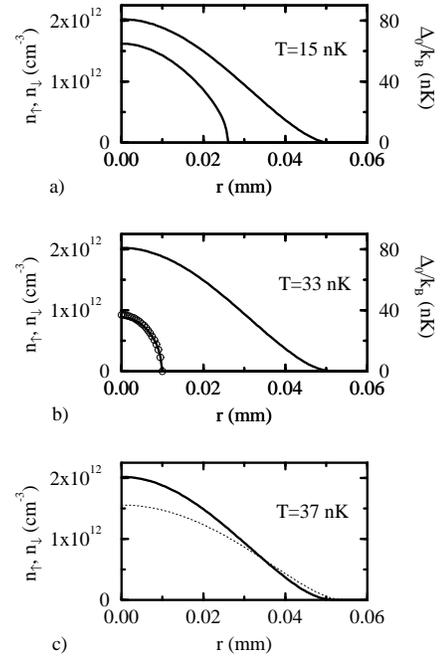}
\caption{\narrowtext 
Density distribution $n_{\uparrow}({\bf r})=n_{\downarrow}({\bf r})$\ and energy
gap $\Delta_0({\bf r})$\ for a $^6$Li atomic gas consisting of $2.865\times 10^5$\
atoms in each spin state at $a)$ $T=15$nK,
$b)$ at $T=33$nK, slightly below $T_c$, and $c)$ at $T=T_c=37$nK. The left
scale of each plot refers to the density, the right scale to energy gap. 
The open circles in $b)$\ represent Eq.~(\ref{deltat}), and the dotted line
in $c)$\ shows the density distribution for a gas with the same number of
particles and at the same temperature, but with $a=0$.} 
\label{fig9}
\end{figure}

In Fig.~\ref{fig9} the density profile $n_{\uparrow}({\bf r})=
n_{\downarrow}({\bf r})$\ and the energy gap $\Delta_0({\bf r})$\ is plotted
for several temperatures below and at the critical temperature
for a gas with $N_{\uparrow} = N_{\downarrow} = 2.865\times 10^5$\ 
particles in both hyperfine states. The dotted line in
Fig~\ref{fig9}$c$\ shows the density distribution for a gas with the
same number of particles, but with $a=0$\ instead of $a=-2160a_0$. It
is clearly visible that the effect of the interaction on the
density is rather large. Indeed, due to the attractive interactions
the particles are pulled to the center of the trap, and the density is
there considerably increased which is good from an experimental point of
view because it significantly increases the critical temperature. 

\begin{figure}[htbp]
\psfig{figure=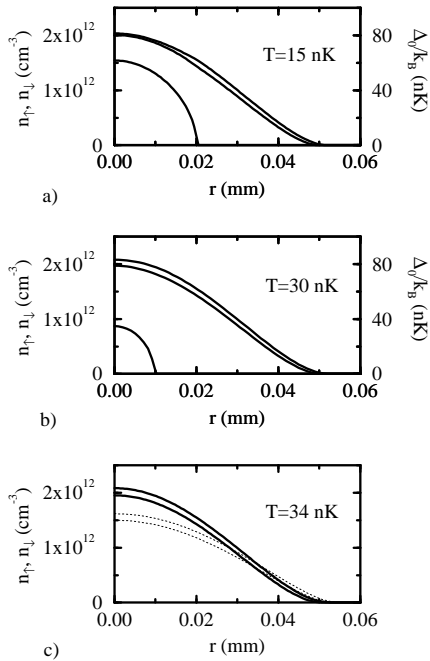}
\caption{\narrowtext 
Density distribution $n_{\uparrow}({\bf r})$, $n_{\downarrow}({\bf r})$\ 
and energy
gap $\Delta_0({\bf r})$\ for a $^6$Li atomic gas with $N_{\uparrow}=3.08\times
10^5$\ and $N_{\downarrow}=2.65\times 10^5$\ below and at the critical 
temperature. In $a)$\ $T=15$ nK,
$b)$ at $T=30$ nK, slightly below $T_c$, and $c)$ at $T=T_c=34$ nK. The left
scale of each plot refers to the density, the right scale to energy gap. The
dotted lines in $c$) represent the density profiles for a non-interacting
gas, with the same number of particles $N_{\uparrow}=3.08\times 10^5$, and
$N_{\downarrow}=2.65\times 10^5$, respectively.} 
\label{fig10}
\end{figure}

Fig.~\ref{fig10} is a similar plot, but now with unequal number
of particles in each spin state: $N_{\uparrow}=3.08\times 10^5$, and
$N_{\downarrow}=2.65\times 10^5$, so the total number of particles 
is the same as in the previous case.  
From Fig.~\ref{fig10}$a$ it can be seen that the presence 
of the order parameter tends to decrease the difference in densities
of each hyperfine level. This can physically 
be understood from the fact that the most favorable condition
for the formation of Cooper pairs is that both densities are equal. 

The most important observation that we can make
from both Figs.~\ref{fig9} and \ref{fig10}, 
is that there is almost no change in the 
density of the gas going from the normal to the superfluid phase. This also
leads, as will be explained in more detail below, to
the conclusion that a measurement of collective excitations
will not give a good signature for the presence of a superfluid state 
\cite{leggett2}.
A second observation is that from the BCS theory in superconductors \cite{degennes},
it is known
 that the order parameter $\Delta_0$\ close to the critical temperature
vanishes as 
\begin{equation}
\Delta_0(T) \simeq 1.74 \Delta_0(0) \sqrt{1-T/T_c}, 
\label{deltat}
\end{equation}
where $\Delta_0(0)$ is the zero-temperature value of $\Delta_0$, which in
turn is related to the critical temperature as 
\begin{equation}
\Delta_0(0) \simeq 1.76 k_BT_c. 
\label{delta0}
\end{equation}
For Fig.~\ref{fig9}$a$\ it follows from Eq.~(\ref{tc}) that the critical
temperature corresponding to the density of the gas in the center of the
trap is much larger than the temperature ($T=15$ nK) itself. Hence the value of the
order parameter approaches the zero-temperature limit in this case. Using that
$T_c[n({\bf 0})]  \simeq 37$nK, one finds from Eq.~(\ref{delta0}) that
$\Delta_0(0)/k_B = 65.1$nK. Comparing this with $\Delta_0({\bf r}={\bf 0})/k_B
=65.0$nK, we find that there is indeed a rather good agreement.
Also, since the temperature in Fig.~\ref{fig9}$b$\ is only slightly below
the critical temperature, we can compare relation Eq.~(\ref{deltat})
with the values of $\Delta_0({\bf r})$\ in this figure. At each spatial
position where $\Delta_0({\bf r}) > 0$, we can from the local value of
the Fermi energy $\epsilon_F({\bf r})$, extract the local critical temperature
$T_c({\bf r})$ from Eq.~(\ref{tc}) and use Eq.~(\ref{delta0}) to compare 
$\Delta_0(T[{\bf r}])/k_B=3.06 T_c({\bf r})\sqrt{1-T/T_c({\bf r})}$\ 
(open circles in Fig.~\ref{fig9}$b$) with $\Delta_0({\bf r})$. Again, the
agreement is very good. 

Finally, we want to check that our local density approximation is indeed
valid under the conditions of interest.
From Fig.~\ref{fig9}, one finds that at $r=0$, the value of $1/k_Fl \simeq
0.06$, and for example at $r=0.05$ mm, we find $1/k_F l\simeq 0.64$. 
Furthermore,
from the zero-temperature value of $\Delta_0$ in Eq.~(\ref{delta0}), we
find that the size of the Cooper pairs relative to the trapping
parameter $l$ is about $0.58$, so the local density approximation
starts to break down if we are far below the critical temperature. In that
case a more accurate approach is required, at least for the relatively
large trapping frequencies used here.

\section{DISCUSSION AND CONCLUSION}
\label{conclusie}

As mentioned above, an important experimental problem is the detection of the 
superfluid state. In contrast to the Bose-Einstein condensation experiments,
there is no clear signature in the density distribution when
the gas becomes superfluid, as shown in 
Sec.~\ref{benedenTc}. Therefore, a measurement
of the collective excitations, or density fluctuations, will not
provide useful information on the presence of the superfluid phase as well.   
This can also be understood from the dissipationless (linear)
hydrodynamic equations governing the density fluctuations in
the system. Considering only the optimal situation of an equal number
of particles $N_{\uparrow}=N_{\downarrow}$, 
these equations are given, for a gas trapped in an external potential 
$V({\bf r})$, by
\begin{mathletters}
\begin{eqnarray}
\label{dndt}
\frac{\partial n}{\partial t} + {\bf \nabla}\cdot {\bf j}^{n} & = & 0 \\
\label{dgdt}
\frac{\partial {\bf j}^n}{\partial t} + \frac{1}{m}\left( 
{\bf \nabla} p + n {\bf \nabla} V({\bf r}) \right)& = & 0 \\
\label{dedt}
\frac{\partial \varepsilon}{\partial t} + {\bf \nabla} \cdot {\bf j}^{\varepsilon}
& = & 0 \\
\label{dvsdt}
\frac{\partial {\bf v}_s}{\partial t} + \frac{1}{m} {\bf \nabla}\mu & = & 0,
\end{eqnarray}
\label{hydro}
\end{mathletters}

\noindent
where $n=n_n+n_s$ is the total density that consists of a normal and superfluid
part, ${\bf j}^n= n_n {\bf v}_n + n_s {\bf v}_s$ is the density current with
${\bf v}_s$ (${\bf v}_n$) the superfluid (normal) velocity, 
$\varepsilon$ the
average energy density, ${\bf j}^{\varepsilon}= \mu {\bf j}^n + Ts{\bf v}_n$ 
the energy current, and $s$ is the entropy density \cite{nozieres,mh2}. 
Note that the same equations
in fact also describe the collective modes of a trapped Bose-condensed gas
\cite{stringari,griffin}.

To show that these hydrodynamic equations result in identical equations
for the collective excitations in the normal and the superfluid phase,
we first of all note that for the densities of interest 
the gas will be in the hydrodynamic limit, meaning that
the time scales for the density fluctuations (which are of the order
of the inverse trapping frequency) are
much slower than the time between elastic collisions. 
In this hydrodynamc regime, density fluctuations and temperature fluctuations
influence each other with a coupling proportional to $c_V-c_p$, 
where $c_V$ ($c_p$) is the heat 
capacity per particle at constant volume (pressure). However, for 
the very low temperatures of interest, one can assume that the heat capacities
of the gas satisfy $c_V \simeq c_p$. Indeed, for a homogeneous Fermi
gas, one finds in the limit $T\downarrow 0$ that
$(c_p -c_V)/c_V = {\cal O} ([k_B T/\epsilon_F]^2)$ and is thus very small.
As a result, the density and temperature fluctuations are effectively uncoupled
\cite{nozieres}. As a consequence, Eq.~(\ref{dedt}), descibing second sound,
decouples from the other 3 equations,
and it suffices to consider only density fluctuations 
$n({\bf r},t)=n_0({\bf r}) + \delta
n({\bf r},t)$. Note also that if we have an unequal number of particles, i.e.\
$N_{\uparrow} \neq N_{\downarrow}$, the density fluctuations are coupled to
fluctuations in the `magnetization' $n_{\uparrow} - n_{\downarrow}$ and
we need to generalize these equations. For equal number of particles these
`spin waves' decouple, however, as we have seen in Sec.~\ref{stability}. 

In the normal phase, 
the Josephson relation Eq.~(\ref{dvsdt}) must be dropped.
Linearizing Eq.~(\ref{dgdt}), which is in fact just Newton's law, 
we arrive at
\begin{eqnarray*}
n_0({\bf r}) \frac{\partial {\bf v}({\bf r},t)}{\partial t} & = &- \frac{1}{m}
{\bf \nabla} \left(
p[n_0({\bf r})] + \frac{\partial p}{\partial n} \delta n({\bf r},t) 
\right)  \\
& & \hspace{0.5cm} - \frac{1}{m}
n_0({\bf r}) {\bf \nabla}V({\bf r}) - \frac{1}{m} \delta n({\bf r},t) 
{\bf \nabla}V({\bf r})  \\
& = & -\frac{1}{m}
{\bf \nabla}\left( \frac{\partial p}{\partial n} \delta n({\bf r},t) \right)
-\frac{1}{m} \delta n({\bf r},t) {\bf \nabla} V({\bf r}),
\end{eqnarray*}
where in the second line we used that in equilibrium 
\begin{equation}
{\bf \nabla} p[n_0({\bf r})] = - n_0({\bf r}) {\bf \nabla} V({\bf r})
\label{EW1}
\end{equation}
and furthermore that the pressure $p$ is a function of the density only at
fixed temperature. This result, together with the continuity equation 
in Eq.~(\ref{dndt}) describes first sound in a trapped Fermi gas. 

In the superfluid phase, first sound has ${\bf v}_s = {\bf v}_n$ and
Eq.~(\ref{dgdt}) now becomes 
\begin{eqnarray}
n_0({\bf r}) \frac{\partial {\bf v}_s}{\partial t} & = & -\frac{1}{m}
{\bf \nabla} \left(
\frac{\partial p}{\partial n} \delta n({\bf r},t) \right) 
- \frac{1}{m} \delta n({\bf r},t) {\bf \nabla}V({\bf r}) \nonumber \\
& = & -\frac{1}{m} \left( {\bf \nabla} \frac{\partial p}{\partial n} +
{\bf \nabla} V({\bf r}) \right) \delta n({\bf r},t)  \nonumber \\
& & \hspace{0.5cm} - \frac{1}{m} \frac{\partial p}{\partial n} 
{\bf \nabla} \delta n({\bf r},t).
\label{dgdtonder}
\end{eqnarray}
However, in the superfluid phase we also have the Josephson relation
Eq.~(\ref{dvsdt}) for the
superfluid velocity. Using that the local chemical potential $\mu = 
\mu_0[n_0({\bf r})
+ \delta n({\bf r},t)] + V({\bf r})$, 
where $\mu_0$ is the homogeneous chemical potential
including the effects of interactions, and linearizing also this equation leads to
\[
n_0({\bf r}) \frac{\partial {\bf v}_s}{\partial t}  =  -\frac{n_0({\bf r})}{m}
{\bf \nabla} \left( \mu_0[n_0({\bf r})] + \frac{\partial \mu_0}{\partial n}
\delta n({\bf r},t) + V({\bf r}) \right)
\]
\begin{equation} 
\hspace{0.5cm} 
= -\frac{n_0({\bf r})}{m} \left( \delta n({\bf r},t) {\bf \nabla} 
\frac{\partial \mu_0}{\partial n}
+ \frac{\partial \mu_0}{\partial n} {\bf \nabla}
\delta n({\bf r},t) \right)
\label{dvsdtonder}
\end{equation}
because in equilibrium the chemical potential must satisfy 
\begin{equation}
\mu_0[n_0({\bf r})] + V({\bf r}) = \mbox{constant}.
\label{EW2}
\end{equation}
Since in general 
\begin{equation} 
\frac{\partial p}{\partial n} = n_0 \frac{\partial \mu_0}{\partial n},
\label{dpdn}
\end{equation}
the second term on the right-hand side of Eq.~(\ref{dvsdtonder}) equals the
last term on the right-hand side of Eq.~(\ref{dgdtonder}). Moreover,
the first term on the right-hand side of Eq.~(\ref{dvsdtonder}) can be
rewritten with Eq.~(\ref{dpdn}) as 
\begin{eqnarray*}
{\bf \nabla} \frac{\partial \mu_0}{\partial n}   
& = & -\frac{\partial p}{\partial n}
\frac{1}{n_0^2({\bf r})} {\bf \nabla} n_0({\bf r}) + \frac{1}{n_0({\bf r})} 
{\bf \nabla} \frac{\partial p}{\partial n}  \\
& = & -\frac{1}{n_0^2({\bf r})} {\bf \nabla }p + \frac{1}{n_0({\bf r})} 
{\bf \nabla} \frac{\partial p}{\partial n}  \\
& = & \frac{1}{n_0({\bf r})} \left( {\bf \nabla} \frac{\partial p}{\partial n} +
{\bf \nabla} V({\bf r}) \right), 
\end{eqnarray*}
where we used again the equilibrium condition in Eq.~(\ref{EW1}). We thus find that
below the critical temperature, the Josephson relation Eq.~(\ref{dvsdt}) 
is identical to the momentum equation (\ref{dgdt}). As in
the normal phase, first sound can below $T_c$ thus be described
merely by Eqs.~(\ref{dndt}) and (\ref{dgdt}). In combination
with the results that the density profiles in the normal and superfluid phase
are almost equal, we conclude that the hydrodynamic equations that describe
the density fluctuations are almost identical, and therefore, that
there will be no significant difference in the
collective excitation spectrum in the superfluid and normal phase respectively.
Consequently, other means of experimental detection must be investigated. 
Of course, this conclusion is based on experiments that couple directly to 
density fluctuations such as in Refs.~\cite{JILA3,MIT3}. If one can couple also
to second sound one would of course observe an additional mode below $T_c$.

Another possible way to detect the superfluid state is by a measurement of the
two-body decay rate of the gas. Note that, in our case,
three-body recombination processes are strongly suppressed, since we are
dealing with fermions, and only have two different hyperfine states
occupied. Above the critical temperature, the two-body rate constants
are essentially
independent of $T$, and the magnitude is depicted as a function of the
applied magnetic field in Fig.~\ref{fig2}. In analogy to the case
of a Bose gas, where the presence of a condensate decreases the decay rate 
due to two-body processes by a factor of about 2 \cite{kagan,stoof3,eric},     
we now analyze the change in the decay rate due to the presence of Cooper pairs 
in the Fermi gas below the critical temperature. 

Using the correlator method from Ref.~\cite{kagan}, it is found that
the decay rate constant due to two-body processes is given by
\begin{equation}
G(T)=G(T_c) K^{(2)}(T),
\label{rate}
\end{equation}
where the correlator 
\begin{equation}
K^{(2)}(T) = \frac{1}{n_{\uparrow}n_{\downarrow}}
\langle \psi^{\dag}_{\uparrow} ({\bf x}) 
\psi^{\dag}_{\downarrow} ({\bf x}) \psi_{\downarrow}({\bf x}) 
\psi_{\uparrow}({\bf x}) \rangle
\label{correlator}
\end{equation} 
equals 1 above $T_c$, but
increases due to the nonzero expectation value $\langle  
\psi_{\uparrow}({\bf x}) \psi_{\downarrow}({\bf x}) \rangle$\ below
the critical temperature.
Indeed, using the transformations given by Eqs.~(\ref{secondq}) and 
(\ref{Bogtrans}), it is found that 
\[
K^{(2)}(T) = 1 + \frac{|\Delta_0|^{2}}{V_{0}^2 n_{\uparrow} n_{\downarrow}},
\]
where $\Delta_0/V_0$\ is again given by the ultra-violet diverging
expression in Eq.~(\ref{gapeq1}).
The question now arises what we should use for $V_0$ in this expression.
Clearly, the denominator $V_0$ should not be considered as the zero-momentum 
component of the triplet potential. Physically, this 
can be understood by the fact that this value of $V_0$\ does not characterize
the exact non-local two-body triplet potential in any way. It does not even
reproduce the correct long distance behavior for the scattered wave function. 
Therefore, a first guess
would be to replace $V_0$ by $T^{2B}$, but since the $s$-wave scattering
length in this case is much larger than the effective range $r_V$\ of the
interaction $V_T({\bf x})$, it is likely that the replacement of $V_0$ by
$T^{2B}$\ underestimates the effect of the presence of the superfluid
phase on the decay rates considerably. 

Instead, from our procedure to remove the ultra-violet divergence in the
gap equation, we see that $V_0$ should be chosen such that it satisfies
the Lippmann-Schwinger equation 
\[
\frac{1}{T^{2B}} = \frac{1}{V_0} + \frac{1}{V} \sum_{|{\bf k|} 
\leq \Lambda} \frac{1}{2 \xi_{\bf k}},
\]
where a cut-off $\Lambda= {\cal O}(1/r_V)$ is introduced.
Solving this equation for $V_0$, and using that
the Fermi energy $\epsilon_F \ll \hbar^2 \Lambda^2/2m$,
we find that
\begin{equation}
V_{0} = T^{2B} \frac{1}{1 - \frac{2a\Lambda}{\pi}}.
\label{Vrenorm}
\end{equation}

Applying this result we can now make an estimate
of the effect of the presence of the superfluid phase.
To simplify the calculation, we will calculate the effect at $T=0$\ for
a homogeneous gas with $n_{\uparrow} = n_{\downarrow} =n$. We then have that 
$n=k_F^3/(6\pi^2)$. Moreover, from the BCS theory we know that
the zero-temperature value of the order parameter $\Delta_0(0)$\ is
related to the critical temperature by $\Delta_0(0) = 1.76 k_B T_c$. So 
\[
\frac{1}{n_{\uparrow}n_{\downarrow}} \frac{|\Delta_0|^2}{V_0^2} 
 \simeq  \frac{1}{n^2} \left( \left[ 1-\frac{2a\Lambda}{\pi} \right] 
\frac{1.76 k_B T_c}{\epsilon_F} \frac{3 \pi n}{4 k_Fa} \right)^2.
\]
From the functional behavior of the triplet potential $V_T({\bf x})$, 
we deduce that the range of the two-body interaction $r_V \simeq 100a_0$. 
Therefore, substituting $\Lambda \simeq (100a_0)^{-1}$ and
$a = -2160 a_0$, we find that in the case of $^6$Li atoms, 
$V_{0} \simeq 0.07 T^{2B}$. Note that, as expected,
the Fermi wavenumber $k_F$\ is much smaller than the cut-off value $\Lambda$:
For a density $n_{\uparrow} = n_{\downarrow} = 10^{12}$ cm$^{-3}$,
we have $\epsilon_F \simeq 6\times 10^{-7} k_B$, resulting in $k_F \simeq
(5\times 10^3 a_0)^{-1}$, which is indeed much smaller than the cut-off $\Lambda$.
Using that $T_c=11$ nK for $n=10^{12}$ cm$^{-3}$ (see Fig.~\ref{fig4}) 
it turns out that $1-2a\Lambda/\pi \simeq 15$ and the 
correlator $K^{(2)}(0) \simeq 7$, which is much larger than its
value above the critical temperature. Had we used $T^{2B}$ instead of $V_0$,
the change in the correlator would have been only of the order of $3\%$.
Even though we do not expect that the corrections to the decay rates are as 
large as the above crude argument suggests, 
we do believe that it might be of the order unity, and may be
measurable.  
Of course, the cut-off dependence of $V_0$\ should in some way drop
out of the theory eventually, but for this a better theory is
needed, which takes into account the precise details of the triplet 
potential and does not make use of a pseudo-potential to replace it.
Work along these lines is in progress, because of the experimental
importance to have a reliable estimate of the changes in the relaxation
rate constants.
Furthermore, note that the correlator $K^{(2)}(T)$ from 
Eq.(\ref{correlator}) also appears in the expression of the average energy
of the system \cite{MIT2}. A measurement of this quantity has been done
for the case of Bose gases \cite{MIT4,JILA4}, 
and we believe that also in the case of
fermions, a change in the average energy at the critical temperature 
can signal the presence of the superfluid phase.

In summary, we considered a gas of atomic $^6$Li occupying two hyperfine
states trapped in a magnetic field. Atoms in different hyperfine levels
can interact via $s$-wave scattering. Using the most up-to-date triplet
potential, we showed that the lifetime
of such a gas with a density of $10^{12}$ cm$^{-3}$ per hyperfine level
is of the order of 1 second, when a magnetic bias field of
5 T is applied. At this density the gas 
becomes superfluid at a temperature of about 11 nK. 
We also investigated the mechanical stability of a two-component
Fermi gas, and showed that, if the two-particle interaction is repulsive,
the gas is unstable for spin-density fluctuations, whereas in case the
interatomic interaction is attractive, the gas is unstable against density
fluctuations above the spinodal line.

Furthermore we considered the superfluid state of atomic $^6$Li,  
trapped in an isotropic harmonic oscillator potential, in the
local density approximation. We showed that the critical temperature
$a/\lambda_{T_c}$ is a universal function of the quantities
$N_{\uparrow}^{1/6}a/l$ and $N_{\downarrow}^{1/6}a/l$.    
Below the critical temperature there is almost no change in the
density profile of the gas cloud. Therefore we suggest that the presence
of the superfluid state might be signaled by a change in the decay rates
or a change in the average energy at the critical temperature.

\section*{ACKNOWLEDGMENTS}
We acknowledge various useful discussions with Yvan Castin,
Tony Leggett, Andrei Ruckenstein, and Peter Zoller. The work in Utrecht
was supported by the Stich\-ting Fundamenteel Onderzoek der
Materie (FOM) which is financially supported by the Nederlandse
Organisatie voor Wetenschappelijk Onderzoek (NWO).
The work at Rice was supported by the National Science Foundation, NASA, the
Texas Advanced Technology Program, and the Welch Foundation.
 

\end{multicols}
\end{document}